\documentclass[reprint,a4paper,aps,pra,amsmath,amsfonts,floatfix,nofootinbib]{revtex4-2}
\usepackage{bm}                      
\usepackage{graphicx}                
\usepackage{dcolumn}                 
\usepackage[table]{xcolor}
\usepackage{hyperref}
\hypersetup{
colorlinks=true,
linkcolor=blue,
citecolor=blue,
citebordercolor=blue,
}

\newcommand*{\cent}[1]{\multicolumn{1}{c}{$#1$}}
\newcolumntype{x}[1]{D{.}{.}{#1}}
\newcommand{\ds}{\displaystyle}
\newcommand{\nn}{\nonumber}
\newcommand{\icm}{\text{cm}^{-1}}
\newcommand{\br}{\vec{r}}

\newcommand{\intV}{\int\!dV}
\newcommand{\riA}{r_{1A}}
\newcommand{\rjA}{r_{2A}}
\newcommand{\riB}{r_{1B}}
\newcommand{\rjB}{r_{2B}}
\newcommand{\rij}{r_{12}}
\newcommand{\oh}{\frac{1}{2}}
\newcommand{\Li}{\mathrm{Li}_2}

\newcommand{\GAB}{G_{AB}}
\newcommand{\Gij}{G_{12}}
\newcommand{\GiB}{G_{1B}}

\newcommand{\pdop}[2]{\left(\frac{\partial}{\partial {#1}}\right)^{#2}}
\newcommand{\pdopd}[2]{\frac{\partial {#1}}{\partial {#2}}}
\newcommand{\hO}{\hat\Omega}
\allowdisplaybreaks

\begin{document}

\title{
Integrals for relativistic nonadiabatic energies of H\texorpdfstring{$_2$}{2} in exponential basis}

\author{Krzysztof Pachucki}
\email{krp@fuw.edu.pl}
\affiliation{Faculty of Physics,
             University of Warsaw, Pasteura 5, 02-093 Warsaw, Poland}

\author{Jacek Komasa} 
\email{komasa@man.poznan.pl}
\affiliation{Faculty of Chemistry, Adam Mickiewicz University,
             Uniwersytetu Poznańskiego 8, 61-614 Pozna{\'n}, Poland}

\date{\today}

\begin{abstract}
Accurate predictions for hydrogen molecular levels require the treatment of electrons and nuclei on an equal footing. While nonrelativistic theory has been effectively formulated this way, calculation of relativistic and quantum electrodynamic effects using an exponential basis with explicit correlations that ensure well-controlled numerical precision is much more challenging. In this work, we derive a complete set of integrals for the relativistic correction and demonstrate their application to several of the lowest rovibrational levels. Together with similar advancements for quantum electrodynamic corrections, this will improve the accuracy beyond $10^{-9}$ and hopefully explain discrepancies with recent experimental values.
\end{abstract}

\maketitle

\section{Introduction}

Molecular hydrogen is the most abundant molecule in the Universe \cite{Fraser:02}. 
It is also the dominant component of the atmosphere of giant planets in the Solar System \cite{Margolis:73}.
Hence, it draws the attention of astronomers and laboratory physicists 
\cite{Roueff:19,Tchernyshyov:22,Tan:22,Gordon:22}. In particular, laboratory spectroscopy provides indispensable data for, e.g., constructing astronomical models and databases 
\cite{Sung:23,Ochsenbein:00,Tennyson:20,Wcislo:16a}, 
determining physical constants \cite{Puchalski:20,Puchalski:22}, 
or searching for new physics beyond the Standard Model \cite{Salumbides:13,Ubachs:16,Hollik:20}. 
In recent years, precision spectroscopy of molecular hydrogen has reached an accuracy
that enables testing the quantum electrodynamic (QED) theory at an accuracy level of several ppb
\cite{Altmann:18,Cheng:18,Beyer:19,Fleurbaey:23,Cozijn:23,Lamperti:23}.

In several recent studies, a systematic discrepancy on the level of 1.5-2.0 MHz ($\sim$5-7$\cdot 10^{-5}\,\icm$) between theoretical and experimental vibrational transition energies of H$_2$, HD, and D$_2$ has been reported  \cite{Fast:20,Lamperti:23,Fleurbaey:23,Cozijn:23}. This inconsistency corresponds to $1$-$3\,\sigma$ of theoretical uncertainty. As an illustration, we can quote the currently most accurate experimental energy for the $S_2(0)$ rovibrational transition in H$_2$: $252\,016\,361.164(8)$ MHz \cite{Cozijn:23}. The corresponding theoretical prediction is $252\,016\,358.6(16)$ MHz \cite{H2Spectre} and differs from the measured value by 2.6 MHz, i.e., 1.6 $\sigma$.  Given that the theoretical nonrelativistic energy is known with kHz ($\sim10^{-7}\,\icm$) accuracy, incomplete accounting for nuclear motion in relativistic and/or QED components of the total energy is most likely the source of these discrepancies. 

In this study, we tackle relativistic correction by treating electrons and nuclei on an equal footing.
We introduce a computational method that achieves an accuracy of a few kHz, similar to that 
for nonrelativistic energies.
We employ the nonadiabatic James-Coolidge (naJC) basis function, which has previously been used 
to solve the four-body Schr{\"o}dinger equation \cite{PK:16} yielding the nonrelativistic energy 
of rovibrational levels with a relative accuracy of $10^{-13}$-$10^{-14}$ \cite{PK:18a,PK:18b,PK:19,PK:22}.
This approach retains its accuracy for rotationally and vibrationally excited states.
Additionally, this accuracy surpasses the uncertainty arising from the imprecise nuclear masses.
The naJC wavefunction fully accounts simultaneously for both the electron correlation 
and the movement of the nuclei. This means that there is no need to separate the electronic 
and nuclear degrees of freedom, nor introduce common approximations such as the one-electron 
or the Born-Oppenheimer. Evaluation of matrix elements with the nonrelativistic
Hamiltonian necessitated finding a new class of integrals, which was the main difficulty
in constructing the naJC-based method.

Applying the naJC wavefunction to relativistic and QED corrections is even more involved. 
Matrix elements of the relativistic Breit-Pauli Hamiltonian in the basis of naJC functions require 
the evaluation of thus-far-unknown classes of integrals. Determination of these integrals is 
the {\em sine qua non} of developing this new approach and of achieving the accuracy needed 
for testing QED. 
This paper presents methods and techniques employed in the evaluation of three new classes 
of such relativistic integrals and presents a proof of concept for the lowest rovibrational 
levels of H$_2$.

\subsection{Wavefunction}

The nonadiabatic James-Coolidge basis function is a special case of a general four-particle exponential
function of the form
\begin{widetext}
\begin{align}\label{Eq:gef}
\psi(\br_1,\br_2,\br_A,\br_B;\{w_j,u_j,n_i\})
&=e^{-w_1\,\rij-w_2\,\rjA-w_3\,\rjB-u_1\,r_{AB}-u_2\,\riB-u_3\,\riA} \nn\\
&\quad\times r_{AB}^{n_0}\,\rij^{n_1}\,(\riA-\riB)^{n_2}\,(\rjA-\rjB)^{n_3}\,(\riA+\riB)^{n_4}\,(\rjA+\rjB)^{n_5}.
\end{align}
\end{widetext}
This function contains all inter-particle distances $r_{ij}=|\br_i-\br_j|$ with $\br_1,\,\br_2$
pointing at electrons and $\br_A,\,\br_B$ -- at nuclei. The non-linear parameters $w_j$ and $u_j$
are assumed to be positive real numbers, and the exponents $n_i$ are non-negative integers.
Matrix elements of the nonrelativistic Hamiltonian evaluated with
these general exponential functions lead to integrals of the form
\begin{widetext}
\begin{align}
g(w_1,w_2,w_3,u_1,u_2,u_3,\{n_i\})
&= \int \frac{d^3 r_{12}}{4\,\pi}\,\int \frac{d^3 r_{2A}}{4\,\pi}\,\int \frac{d^3 r_{2B}}{4\,\pi}\,
\frac{e^{-w_1\,r_{12}}}{r_{12}}\,\frac{e^{-w_2\,r_{2A}}}{r_{2A}}\,
\frac{e^{-w_3\,r_{2B}}}{r_{2B}}\,\frac{e^{-u_1\,r_{AB}}}{r_{AB}}\,
\frac{e^{-u_2\,r_{1B}}}{r_{1B}}\,\frac{e^{-u_3\,r_{1A}}}{r_{1A}}\nn\\
&\quad\times 
 r_{AB}^{n_0}\,\rij^{n_1}\,(\riA-\riB)^{n_2}\,(\rjA-\rjB)^{n_3}\,(\riA+\riB)^{n_4}\,(\rjA+\rjB)^{n_5}\label{Eq:gnon}.
\end{align}
\end{widetext}
The sequence of integer exponents $n_0,n_1,n_2,n_3,n_4,n_5$ is denoted as $\{n_i\}$. 
When this symbol is omitted, it means $\{0\}\equiv 0,0,0,0,0,0$, and the corresponding integral 
is called the master integral.
It is convenient to express the function~(\ref{Eq:gef}) in elliptic-like variables:
\begin{align}\label{Eq:ellcoord}
\zeta_1&=\riA+\riB\,, \quad \eta_1=\riA-\riB\,, \quad \zeta_2=\rjA+\rjB\,, \nn\\ \eta_2&=\rjA-\rjB\,, \quad R=r_{AB}\,,
\end{align}
which entails introducing new symbols for linear combinations of parameters
\begin{align}\label{Eq:tw1xyuw}
&w_2=w+x,\qquad w_3=w-x,\qquad u_2=u-y,\nn\\ 
&u_3=u+y,\qquad u_1=t.
\end{align}
In this notation
\begin{align}\label{Eq:gefell}
&\psi(\br_1,\br_2,\br_A,\br_B;t,w_1,y,x,u,w,\{n_i\})\\
&=e^{-t\,R-w_1\,\rij-y\,\eta_1-x\,\eta_2-u\,\zeta_1-w\,\zeta_2}\,
R^{n_0}\,\rij^{n_1}\,\eta_1^{n_2}\,\eta_2^{n_3}\,\zeta_1^{n_4}\,\zeta_2^{n_5},\nn
\end{align}
and corresponding integrals assume the following form
\begin{align}\label{Eq:gnonell}
&g(t,w_1,y,x,u,w,\{n_i\})\nn\\
&\quad= \intV\,\frac{
e^{-t\,R}\,e^{-w_1\,r_{12}}\,e^{-y\,\eta_1}\,e^{-x\,\eta_2}\,e^{-u\,\zeta_1}\,e^{-w\,\zeta_2}
}{R\,\rij\,\riA\,\riB\,\rjA\,\rjB}\nn\\
&\hspace{11ex}\times
R^{n_0}\,\rij^{n_1}\,\eta_1^{n_2}\,\eta_2^{n_3}\,\zeta_1^{n_4}\,\zeta_2^{n_5},
\end{align}
where we introduced the shorthand symbol
$\ds\intV\equiv\int \frac{d^3 r_{12}}{4\,\pi}\,\int \frac{d^3 r_{2A}}{4\,\pi}\,\int \frac{d^3 r_{2B}}{4\,\pi}$.
Unfortunately, such integrals are difficult to handle  \cite{Fromm:87,Harris:09},
which prompts for a slight simplification of the general function~(\ref{Eq:gef}).
This simplification is achieved by setting 
\begin{align}
w_1=0, \qquad y=0, \qquad x=0, \quad\text{and}\quad w=u.
\end{align}
The corresponding function
\begin{align}\label{Eq:naJC}
&\psi(\br_1,\br_2,\br_A,\br_B;t,u,\{n_i\})\nn\\
&\quad=e^{-t\,R-u\,(\zeta_1+\zeta_2)}\,
R^{n_0}\,\rij^{n_1}\,\eta_1^{n_2}\,\eta_2^{n_3}\,\zeta_1^{n_4}\,\zeta_2^{n_5}
\end{align}
was named the nonadiabatic James-Coolidge (naJC) function for its resemblance to 
the two-electron James-Coolidge function used in clamped nuclei calculations 
with H$_2$ \cite{James:33}. 

\subsection{Integrals in the James-Coolidge basis}

The whole class of integrals appearing in the matrix elements
of the nonrelativistic Hamiltonian in the naJC basis~(\ref{Eq:naJC}) were implemented 
\cite{PK:16}.
Arbitrary $\{n_i\}$ integrals can be formally defined as multiple derivatives with respect
to non-linear parameters present in the general master integral $g(t,w_1,y,x,u,w)$ of Eq.~(\ref{Eq:gnonell})
\begin{widetext}
\begin{align}
G(t,u;\{n_i\})
&=\left(-\frac{\partial}{\partial t}\right)^{n_0}
\left(-\frac{\partial}{\partial w_1}\right)^{n_1}_{w_1=0}
\left(-\frac{\partial}{\partial y}\right)^{n_2}_{y=0}
\left(-\frac{\partial}{\partial x}\right)^{n_3}_{x=0}
\left(-\frac{\partial}{\partial u}\right)^{n_4}_{u=w}
\left(-\frac{\partial}{\partial w}\right)^{n_5}\,g(t,w_1,y,x,u,w).
\end{align}
\end{widetext}

In the above formulas and in what follows henceforth, we use the notation in which
simplified versions of integrals $g$ will be denoted by capital $G$ and will appear in two variants,
one with $w=u$ and the other one with $w\neq u$
\begin{align}
G(t,u;\{n_i\})&\equiv g(w_1=0,y=0,x=0,w=u),\\
G(t,u,w;\{n_i\})&\equiv g(w_1=0,y=0,x=0).
\end{align}
Hence, writing explicitly
\begin{align}
G(t,u;\{n_i\})\label{Eq:Gtun}
&=\intV\,\frac{e^{-t\,R}\,e^{-u\,(\zeta_1+\zeta_2)}}
              {R\,\rij\,\riA\,\riB\,\rjA\,\rjB}\nn\\
&\qquad\times
R^{n_0}\,\rij^{n_1}\,\eta_1^{n_2}\,\eta_2^{n_3}\,\zeta_1^{n_4}\,\zeta_2^{n_5},\\[1em]
G(t,u)\label{Eq:Gtu}
&=\intV\,\frac{e^{-t\,R}\,e^{-u\,(\zeta_1+\zeta_2)}
}{R\,\rij\,\riA\,\riB\,\rjA\,\rjB} .
\end{align}
From now on, we also assume that the condition $t>2\,u$ is satisfied, and formulas for $-2\,u \leq t \leq 2\,u$ are obtained by analytic continuation.

Techniques developed by one of us (K.P.) to evaluate such integrals were described 
in Refs.~\onlinecite{Pachucki:09,Pachucki:12b}.
In short, this approach relies on a set of partial differential equations (PDE)
to which the integrals $g$ are solutions. 
All these PDEs can be written as
\begin{equation}\label{PDE}
\sigma\,\frac{\partial g}{\partial \beta } +
\frac{1}{2}\,\frac{\partial\sigma}{\partial \beta}\,g + P_{\beta} = 0\,, 
\end{equation}
where $\beta$ is one of the six parameters $t$, $w_1$, $y$, $x$, $u$, or $w$,
and $\sigma$ is the following polynomial in these six parameters
\begin{align}\label{Eq:sigma}
\sigma&=w_1^2\,t^4+w_1^2\,(u+w-x-y) (u-w+x-y)\nn\\
&\hspace{11.4ex}\times (u-w-x+y) (u+w+x+y)\nn\\
&\quad +t^2 \Bigl(w_1^4-2\,w_1^2 \left(u^2+w^2+x^2+y^2\right)+16\,u\,w\,x\,y\Bigr)\nn\\
&\quad-16 (u\,y-w\,x) (u\,x-w\,y) (u\,w-x\,y)\,.
\end{align}
Properly manipulating these equations
leads to recurrence relations in all variables, which enables finding arbitrary
nonrelativistic integrals of Eq.~(\ref{Eq:Gtun}). In particular, the explicit formulae
for the master integrals are
\begin{align}~\label{Eq:Gtmi}
G(t,u,w)
&=\frac{1}{4\,u\,w}\biggl[
 \frac{\ln\frac{2\,u\,w}{(t+u+w)(u+w)}}{t+u+w}
 -\frac{\ln\frac{2\,u}{t+u+w}}{t-u+w}\nn\\
&\hspace{12ex}-\frac{\ln\frac{2\,w}{t+u+w}}{t+u-w}
 +\frac{\ln\frac{2\,(u+w)}{t+u+w}}{t-u-w}
 \biggr],\\
G(t,u)
&=\frac{1}{4\,u^2}\biggl[
 \frac{\ln\frac{u}{t+2\,u}}{t+2\,u}-\frac{2\ln\frac{2\,u}{t+2\,u}}{t}
 +\frac{\ln\frac{4\,u}{t+2\,u}}{t-2\,u}
 \biggr].\label{Eq:Gmi}
\end{align}

Similarly, we proceed with the relativistic integrals resulting from 
evaluating matrix elements with the Breit-Pauli Hamiltonian. 
The new relativistic integrals can be divided into three classes
\begin{align}
\GAB(t,u;\{n_i\})
&=\intV\,\frac{e^{-t\,R}\,e^{-u\,(\zeta_1+\zeta_2)}}
              {R^2\,\rij\,\riA\,\riB\,\rjA\,\rjB}\nn\\&\qquad\times
R^{n_0}\,\rij^{n_1}\,\eta_1^{n_2}\,\eta_2^{n_3}\,\zeta_1^{n_4}\,\zeta_2^{n_5}, \label{EhatG}\\
\Gij(t,u;\{n_i\})
&=\intV\,\frac{e^{-t\,R}\,e^{-u\,(\zeta_1+\zeta_2)}}
              {R\,\rij^2\,\riA\,\riB\,\rjA\,\rjB}\nn\\&\qquad\times
R^{n_0}\,\rij^{n_1}\,\eta_1^{n_2}\,\eta_2^{n_3}\,\zeta_1^{n_4}\,\zeta_2^{n_5}, \label{EddotG}\\
\GiB(t,u;\{n_i\})
&=\intV\,\frac{e^{-t\,R}\,e^{-u\,(\zeta_1+\zeta_2)}}
              {R\,\rij\,\riA\,\riB^2\,\rjA\,\rjB}\nn\\&\qquad\times
R^{n_0}\,\rij^{n_1}\,\eta_1^{n_2}\,\eta_2^{n_3}\,\zeta_1^{n_4}\,\zeta_2^{n_5}. \label{EtildeG}
\end{align}
The remaining integrals ($G_{1A}$, $G_{2A}$, and $G_{2B}$) can be obtained by a permutation of variables.
To find the integrals (\ref{EhatG})-(\ref{EtildeG}) from their master integrals,
we need to established recurrence relation for all six indices $n_0,n_1,n_2,n_3,n_4$, and $n_5$.
Each type of these integrals requires a different treatment. Therefore, a separate section 
will be devoted to each of them. In each section, we will first describe the derivation 
of the pertinent master integral and then the recursive relations in all variables.

\section{\texorpdfstring{$\GAB$}{GAB} integrals}

Let us first note that
\begin{align}\label{Eq:GABeqG}
\GAB(t,u;\{n_i\})&=G(t,u;n_0-1,n_1,n_2,n_3,n_4,n_5)\,,
\end{align}
so all the integrals with $n_0\geq 1$ are considered to be known.
What we need are the remaining $\GAB$ integrals with $n_0=0$,
in particular, the master integral
\begin{align}
\GAB(t,u)&=\intV\,
\frac{e^{-t\,R}\,e^{-u\,(\zeta_1+\zeta_2)}}{R^2\,\rij\,\riA\,\riB\,\rjA\,\rjB}.\label{EhatGm}
\end{align}
The $\GAB(t,u)$ master integral can be found analytically by direct integration of $G(t,u)$
over $t$. For this purpose, we first rearrange Eq.~(\ref{Eq:Gmi})
\begin{align}\label{Eq:Gminew}
G(t,u)
&=-\frac{t\ln{2}-2\,u\,\ln\frac{t+2\,u}{2\,u}}{t\,u\,(t-2\,u)\,(t+2\,u)}\,.
\end{align}
Then, relying on the integral
\begin{equation}\label{Eq:trAB}
\frac{e^{-t\,R}}{R^2}=\ds\int_{t}^\infty\!dt \frac{e^{-t\,R}}{R}\,,
\end{equation}
we evaluate
\begin{align}\label{Eq:inttG}
\GAB(t,u)&=\int_t^\infty\!dt\, G(t,u)
\end{align}
and express the result in terms of  dilogarithms ($\Li$), namely
\begin{align}
\GAB(t,u) =&\  \frac{1}{2\,u^2}\biggl[ \oh\,\Li\biggl(\frac{t-2\,u}{t+2\,u}\biggr) - \Li\biggl(\frac{t}{t+2\,u}\biggr) +\frac{\pi^2}{12}\biggr].
\end{align}
The integration in Eq.~(\ref{Eq:inttG}) can also be performed numerically and confronted with
the analytic result to verify its accuracy. 
All other $\GAB(t,u;0,n_1,n_2,n_3,n_4,n_5)$ integrals were evaluated by numerical integration 
with respect to $t$ of corresponding $G(t,u;0,n_1,n_2,n_3,n_4,n_5)$ functions; therefore,
no recurrences are needed in this case.

\section{\texorpdfstring{$\Gij$}{G12} integrals}

\subsection{The \texorpdfstring{$\Gij(t,u)$}{G12(t,u)} master integral}

Let us note that
\begin{align}
\Gij(t,u;\{n_i\})&=G(t,u;n_0,n_1-1,n_2,n_3,n_4,n_5)\,,
\end{align}
which means that all the $\Gij$ integrals with $n_1\geq 1$ are identical to 
the corresponding nonrelativistic integrals $G$.
Of interest to us are the remaining $\Gij$ integrals with $n_1=0$.
We start with the evaluation of the master integral
\begin{align}
\Gij(t,u)&=\intV\,
\frac{e^{-t\,R}\,e^{-u\,(\zeta_1+\zeta_2)}}{R\,\rij^2\,\riA\,\riB\,\rjA\,\rjB}.\label{EddotGm}
\end{align}
Because this integral does not depend explicitly on the $w_1$ parameter (related to 
the $\rij$ variable), the direct integration method
applied to $\GAB$ functions will not work, and a more sophisticated method, described below, must be applied.

\begin{figure}[b]
\includegraphics[scale=0.75]{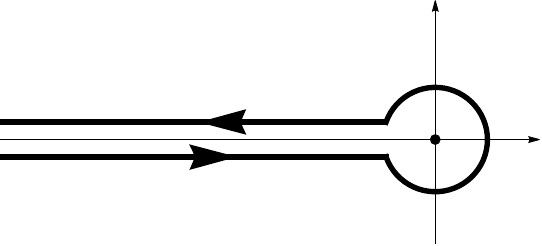}
\caption{Integration path for the Hankel's integral of Eq.~(\ref{Eq:Omega}).}
\label{Fig:Contour}
\end{figure}

Let $g(-w_1) = g(t,w_1,y,x,u,w)$, see Eq.~(\ref{Eq:gnonell}), and consider 
the following Hankel's contour integral (see Eq. (6.1.4) of Ref. 
\onlinecite{Abramowitz:65} and Figure~\ref{Fig:Contour})
\begin{align}
g_\alpha = \hat\Omega\left(\omega^{-\alpha}\,g(\omega)\right) \equiv  \frac{1}{2\,\pi\,i}\,\int_{-\infty}^{(0^+)}g(\omega)\,\omega^{-\alpha}\,d\omega\ \label{Eq:Omega}
\end{align}
for an arbitrary real $\alpha$. 
We show, that, subject to $y=0$, $x=0$, and $w=u$,
\begin{align}
g'_0 \equiv \frac{d g_\alpha}{d \alpha}\biggr|_{\alpha=0} = G_{12}(t,u)\,. \label{Eq:g0pG12}
\end{align}
If in Eq.(\ref{Eq:Omega}) we change the order of $dV$ with $d\omega$ integrations, then the $\omega$-integral takes the form
\begin{align}
\frac{\rij^{\alpha-2}}{\Gamma(\alpha)} = 
\frac{1}{2\,\pi\,i}\,\int_{-\infty}^{(0^+)} \frac{e^{\omega\,\rij}}{\rij}\,\omega^{-\alpha}\,d\omega \,.
\end{align}
Because the derivative at $\alpha=0$ of the left side is
\begin{align}
\frac{d}{d\alpha}\biggr|_{\alpha=0} \frac{\rij^{\alpha-2}}{\Gamma(\alpha)} 
= \frac{d}{d\alpha}\biggr|_{\alpha=0} \alpha\,\rij^{\alpha-2} 
= \frac{1}{\rij^2}\,,
\end{align}
Eq.~(\ref{Eq:g0pG12}) is proved. 

Consider now two PDE's of Eq.~(\ref{PDE}), the first with $\beta=w_1$ and the second with $\beta=t$,
where $\sigma(y=0,x=0,w=u)=w_1^2\,t^2\,(w_1^2+t^2-4u^2)$ and where $P_{w_1}$ and $P_{u_1}$ are taken from Appendix~\ref{App:Pbeta}.
We transform the first equation by substituting $w_1=-\omega$ and multiplying by $t^{-2}\,\omega^{-3-\alpha}$
\begin{align}
&\omega^{-\alpha-1} \left(t^2-4 u^2+\omega ^2\right) \frac{\partial g(\omega)}{\partial \omega }\\
&+\omega^{-\alpha-2} \left(t^2-4 u^2+2 \omega ^2\right) g(\omega) -t^{-2}\,\omega^{-\alpha-3} P_{w_1}(\omega)=0.\nn \label{EPDEw1}
\end{align}
In the next step, we apply $\hO$ of Eq.~(\ref{Eq:Omega}) to the above equation and use the relation
\begin{equation}\label{Eq:Odgtog}
\hO\left(\omega^{-\alpha} \frac{\partial g(\omega)}{\partial \omega }\right)
 =\alpha\,\hO\left(\omega^{-\alpha-1} g(\omega)\right),
\end{equation}
to obtain
\begin{align}
&(\alpha+2)(t^2-4u^2)\,\hO\Bigl(\omega^{-\alpha-2}g(\omega)\Bigr)\\
&+(\alpha+1)\,\hO\Bigl(\omega^{-\alpha}g(\omega)\Bigr)
 -t^{-2}\,\hO\Bigl(\omega^{-\alpha-3} P_{w_1}(\omega)\Bigr)=0\,.\nn
\end{align}
Recalling the definition of the $g_\alpha$ in Eq.~(\ref{Eq:Omega}), we get
\begin{align}
&(\alpha+2)(t^2-4u^2)g_{\alpha+2}+(\alpha+1)g_{\alpha}-G_{w_1}(\alpha+3)=0\,, \label{38}
\end{align}
where
\begin{align}\label{Eq:Gbeta}
G_{\beta}(\alpha) = \frac{1}{t^2}\,\hO\bigl(\omega^{-\alpha} P_{\beta}(\omega)\bigr)\,,
\end{align}
and hence
\begin{align}
g_{\alpha+2}=\frac{-(\alpha+1)g_{\alpha}+G_{w_1}(\alpha+3)}{(\alpha+2)(t^2-4\,u^2)}\,.\label{Eq:ga2}
\end{align}

Now, let us transform the second PDE. Again, we set $w_1=-\omega$, next 
we multiply it by $t^{-1}\,\omega^{-4-\alpha}$, and then apply the $\hO$ operator to get
\begin{align}
&t\,(t^2-4\,u^2)\,\frac{\partial g_{\alpha+2}}{\partial t}
 +t\,\frac{\partial g_\alpha}{\partial t} 
 +2\,(t^2-2\,u^2)\,g_{\alpha+2}+g_\alpha\nn\\
&\qquad = -t\,G_{u_1}(\alpha+4) \,.
\end{align}
Now, we insert $g_{\alpha+2}$ from Eq.~(\ref{Eq:ga2})
and multiply the result by $(2+\alpha)(t^2-4\,u^2)$, obtaining
\begin{align}\label{PDEq:galpha}
&t\,(t^2-4u^2)\frac{\partial g_\alpha}{\partial t}+\left[t^2\,(2+\alpha)-4\,u^2\right]\,g_\alpha=H_\alpha\,,
\end{align}
where
\begin{align}
H_\alpha&= -(2+\alpha)\,t\,(t^2-4\,u^2)\,G_{u_1}(\alpha+4)\\
&\quad-t\,(t^2-4\,u^2)\frac{\partial G_{w_1}(\alpha+3)}{\partial t}+4\,u^2\,G_{w_1}(\alpha+3)\,.\nn
\end{align}
Differentiation of Eq.~(\ref{PDEq:galpha}) with respect to $\alpha$ at $\alpha=0$, bearing in mind that $g_0=0$,
yields the partial differential equation of the form
\begin{align}
&t\,(t^2-4u^2)\frac{\partial g'_0(t)}{\partial t}+\left(2\,t^2-4u^2\right)\,g'_0(t)=H'_0\,.\label{PDEq:g0p}
\end{align}
The solution to this equation is the master integral $g_0'$ we are looking for. First, however, we must find 
an explicit formula for $H'_0$. For this purpose, we evaluate
\begin{align}
H'_0&=\frac{\partial H_\alpha}{\partial \alpha}\biggr|_{\alpha=0} \label{Eq:H0p} \nn\\ 
&= -t\,(t^2-4u^2)\,G_{u_1}(4)-2\,t\,(t^2-4u^2)\,G'_{u_1}(4)\nn\\
&\quad -t\,(t^2-4u^2)\frac{\partial G'_{w_1}(3)}{\partial t}+4u^2\,G'_{w_1}(3)\,.
\end{align}
Explicit formulas for $G_\beta$ functions can be obtained from 
the corresponding $P_\beta$ polynomials, cf. Eq.~(\ref{Eq:Gbeta}), 
and are listed in Appendix~\ref{App:Gbeta}.
After insertion of these formulas to Eq.~(\ref{Eq:H0p}), $H_0'$ simplifies greatly to its final form
\begin{equation}\label{H0p}
H_0'=-\frac{\pi^2}{12}+2\,\Li\left(\frac{t}{t+2\,u}\right).
\end{equation}
We can now return to Eq.~(\ref{PDEq:g0p}) and solve it for $g_0'$.
\begin{align}\label{Eq:g0p}
g_0'=G_{12}(t,u) =\frac{1}{t\,\sqrt{t^2-4\,u^2}}\int_{2u}^t\!dt\,\frac{H_0'}{\sqrt{t^2-4\,u^2}}\,.
\end{align}
The lower integration limit is $2\,u$, because $g'_0$ must be finite at every positive $t$
including $t=2\,u$.

Integration by parts and appropriate variable changes enable the working representation of the above integral,
suitable for effective numerical integration to a desired accuracy
\begin{align}\label{Eq:g0pfinal}
\Gij(t,u)&=
\frac{1}{t\,\sqrt{t^2-4\,u^2}}\Biggl\{H_0'(t,u) \,\ln\left(\tau+\sqrt{\tau^2-1}\right)
\nn\\&\quad
 -\int_0^{\sqrt{\frac{\tau-1}{\tau+1}}}\!dy\,\frac{4\,y}{y^2+1}\,\ln\frac{1-y^2}{2}\,\ln\frac{1-y}{1+y}
 \Biggr\},
\end{align}
where $\tau=t/(2u)$. The numerical integration is performed over a bounded interval with the upper limit 
$\sqrt{\frac{\tau-1}{\tau+1}}<1$, and the integrand is a monotonic function of $y$.
Therefore, the convergence of a numerical quadrature is fast.

\subsection{Recurrences}

Our next goal is to establish recurrence relations which enable the evaluation of an arbitrary integral 
$\Gij(t,u;\{n_i\})$ from integrals with lower values of exponents $n_i$.
We proceed similarly as in the derivation of the master integral. The main difference
is that we set $y=0$, $x=0$, and $w=u$ only after differentiation with $\hat{D}$ of Eq.~(\ref{Eq:D}).
We start by employing the PDE~(\ref{PDE}) with $\beta=w_1$ and $\sigma$ of Eq.~(\ref{Eq:sigma}). 
For clarity, we write the latter as
\begin{align}
\sigma&=w_1^4\,A_{w_1}+w_1^2\,B_{w_1}+C_{w_1} 
\intertext{and}
\frac{1}{2}\,\frac{\partial\sigma}{\partial w_1}&=2\,w_1^3\,A_{w_1}+w_1\,B_{w_1}\,,
\end{align}
where
\begin{align}
A_{w_1}&=t^2,\\
B_{w_1}&=t^4-2\,t^2\left(u^2+w^2+x^2+y^2\right)\nn\\
 &\quad+(u+w-x-y) (u-w+x-y)\nn\\
 &\qquad\times (u-w-x+y) (u+w+x+y),\\
C_{w_1}&=16\,t^2\,u\,w\,x\,y-16\,(u y-w x) (u x-w y) (u w-x y)\,.
\end{align}
Subsequently, we set $w_1=-\omega$, multiply the PDE by $\omega^{-\alpha-3}$, 
apply the operator $\hO$ defined in Eq.~(\ref{Eq:Omega}), and use 
Eqs.~(\ref{Eq:Odgtog}) and~(\ref{Eq:Gbeta}).
As a result, we get
\begin{align}
&(\alpha+1)A_{w_1}\,g_\alpha+(\alpha+2)B_{w_1}\,g_{\alpha+2}+(\alpha+3)C_{w_1}\,g_{\alpha+4}\nn\\
&\quad -t^2\,G_{w_1}(\alpha+3)=0\,.
\end{align}
The obtained equation is differentiated using the following operator
\begin{align}\label{Eq:D}
&\hat{D}\equiv(-1)^{n_2+n_3+n_4+n_5}\\
&\times\pdop{w}{n_5}\bigg|_{w=u}\pdop{u}{n_4}\pdop{x}{n_3}\bigg|_{x=0}\pdop{y}{n_2}\bigg|_{y=0}.\nn
\end{align}
The resulting expression is a long combination of multiple derivatives of 
$g_\alpha$, $g_{\alpha+2}$, and $g_{\alpha+4}$ of the order at most $n_2+n_3+n_4+n_5$
plus a single $G_{w_1}$ term, see Eq.~(\ref{Eq:Gbeta}).
Among them, one $g_\alpha$ and one $g_{\alpha+2}$ function occur with the highest 
{\em shell} of exponents $n_2,n_3,n_4,n_5$. 
Let us extract this $g_{\alpha+2}$ function to obtain the relation
\begin{equation}\label{Eq:gap2}
g_{\alpha+2}(t,u;n_2,n_3,n_4,n_5)= (\dots)\,g_{\alpha}(t,u;n_2,n_3,n_4,n_5) + \ldots
\end{equation}
needed for recursion in the parameter $\alpha$.

Next, we employ another PDE~(\ref{PDE}) with $\beta=y$.
This time, 
\begin{align}
\frac{1}{2}\,\frac{\partial\sigma}{\partial y}=w_1^2\,A_y+B_y\,,
\end{align}
where
\begin{align}
A_{y}&=-2( t^2 y-2 u w x + u^2 y + w^2 y + x^2 y - y^3) ,\\
B_{y}&=8 t^2 u w x + 8 x (-w x + u y) (u x - w y)\nn\\
&\quad + 8 w (-w x + u y) (u w - x y)\nn\\
&\quad - 8 u (u x - w y) (u w - x y)\,.
\end{align}
We treat this PDE in a similar way as the first one; we set $w_1=-\omega$,
multiply it by $\omega^{-\alpha-4}$, and apply the $\hO$ operator, obtaining
\begin{align}\label{Eq:gy}
&A_y\,g_{\alpha+2} + B_y\,g_{\alpha+4}+ A_{w_1}\,\pdopd{g_{\alpha}}{y} + B_{w_1}\,\pdopd{g_{\alpha+2}}{y}\nn\\
&\quad + C_{w_1}\,\pdopd{g_{\alpha+4}}{y} +  t^2\,G_y\,(\alpha+4) = 0\,,
\end{align}
which we differentiate using $\hat{D}$. There are two functions with arguments from the maximal shell,
$g_\alpha(t,u;n_2+1,n_3,n_4,n_5)$ and $g_{\alpha+2}(t,u;n_2+1,n_3,n_4,n_5)$.
We use Eq.~(\ref{Eq:gap2}) to eliminate $g_{\alpha+2}(t,u;n_2+1,n_3,n_4,n_5)$
to get the new relation
\begin{equation}
g_\alpha(t,u;n_2+1,n_3,n_4,n_5)=\dots\,,
\end{equation}
which expresses $g_\alpha$ in terms of the other $g_\alpha$, $g_{\alpha+2}$ and $g_{\alpha+4}$ from lower shells, as well as by
functions $G_{w_1}$ and $G_y$ originating from inhomogeneous terms.

Now, we can repeat this procedure for the other pairs of parameters 
(and pertinent PDEs): 
$(w_1,x)$, $(w_1,u)$, and $(w_1,w)$, each time obtaining the corresponding recursive relation for 
\begin{align}
g_\alpha(t,u;n_2,n_3+1,n_4,n_5)&=\dots,\\
g_\alpha(t,u;n_2,n_3,n_4+1,n_5)&=\dots,\\
g_\alpha(t,u;n_2,n_3,n_4,n_5+1)&=\dots.
\end{align}

From the obtained set of five $g_\alpha$ relations for arbitrary $\alpha$, we get corresponding 
relations at $\alpha=0$ and derivatives in $\alpha$ at $\alpha=0$. The final 10 relationships 
together with the initial $g'_0$ of Eq.~(\ref{Eq:g0p}) form an exhaustive set of recurrences needed to evaluate 
the function $g'_0$ with arbitrary $n_2,n_3,n_4$, and $n_5$
\begin{equation}
g'_0(t,u;n_2,n_3,n_4,n_5)=\Gij(t,u;0,0,n_2,n_3,n_4,n_5)\,.
\end{equation}
Other functions $g_0, g_2, g_4, g'_2, g'_4$ appearing within these relationships are just auxiliary
and serve only to maintain the complete scheme of recurrences.

The last step is to construct $\Gij$ integrals for any exponent $n_0$ from the relation
\begin{align}
&\Gij(t,u;n_0,0,n_2,n_3,n_4,n_5)\nn\\
&\qquad=\left(-\frac{\partial}{\partial t}\right)^{n_0}\Gij(t,u;0,0,n_2,n_3,n_4,n_5)\,.
\end{align}
Inspection of achieved explicit expressions permits the writing of functions to be differentiated in a general form
\begin{align}\label{Eq:G12lc}
&\Gij(t,u;0,0,n_2,n_3,n_4,n_5)\nn\\
&\qquad= u^{-(n_2+n_3+n_4+n_5)}\,\sum_{i=0}^3\,c_i(x)\,f_i
\end{align}
where $x=2\,u/t$,
\begin{align}
f_0&= u^2\,\Gij(t,u), &
f_1&= \frac{\pi ^2}{24}-\text{Li}_2\left(\frac{t}{t+2u}\right), \nn\\
f_2&= \log \left(\frac{2 u}{t+2u}\right), & 
f_3&=1
\end{align}
and where the coefficients $c_i(x)$ are simple rational functions of $x$, for example
\begin{widetext}
\begin{align}  
\Gij(t,u;0,0,0,0,1,0)&= \frac{1}{u^3}\,\frac{2\,f_0 +f_1}{x^2-1} ,\\
\Gij(t,u;0,0,1,1,0,0) &= -\frac{1}{u^4}\,\frac{2\,f_0\,\left(x^2-4\right)
 -f_1\,\left(x^2+2\right)+f_2\,(x-1)\left(x^2-2\right)+(x-1)^2}{(x^2-1)^2}, \\
\Gij(t,u;0,0,0,2,0,0)
&=\frac{x^2}{u^4}\,\frac{f_2\,(x-1)\,x^4-f_1\,(x^2-4)\,x^2+2\,f_0\,\left(x^4-2 x^2+4\right)+(x-1)^2}{16\,(x^2-1)^2}.
\end{align}
\end{widetext}
Differentiating such functions does not pose any particular difficulties.

\section{\texorpdfstring{$\GiB$}{G1B} integrals}

\subsection{The \texorpdfstring{$\GiB(t,u)$}{G1B(t,u)} master integral}

We are going to derive here an explicit formula for the master integral
\begin{align}
\GiB(t,u)&=\intV\,
\frac{e^{-t\,R}\,e^{-u\,(\zeta_1+\zeta_2)}}{R\,\rij\,\riA\,\riB^2\,\rjA\,\rjB}\label{EtildeGm}\,.
\end{align}
For this purpose we express this integral in terms of the derivative of the function $g$
\begin{align}
\GiB(t,u) =&\ \int_{u_2}^\infty\!du_2\,g(t,u_2,u_3,w)\Big|_{u_2=u_3=u}\\
              =&\ -\int_{u}^\infty\!du_2 \int_t^\infty\!dt\, \frac{\partial g(t,u_2,u,w)}{\partial t} \\
              =& -\int_t^\infty\!dt \int_{u}^\infty\!du_2\, \frac{\partial g(t,u_2,u,w)}{\partial t}.
\end{align}
Because $g$ satisfies the PDE~(\ref{PDE}) with $\beta=t=u_1$,
$\sigma=(u_2 - u_3)^2\,(u_2 + u_3)^2\,w^2$,
and $\dfrac{\partial\sigma}{\partial t}=0$,
its derivative can be found immediately
\begin{equation}\label{Eq:dgdt}
\frac{\partial g(t,u_2,u_3,w)}{\partial t } = -\frac{P_{u_1}}{\sigma}\,.
\end{equation}
Hence, we arrive at
\begin{align}
\GiB(t,u) = & -\int_t^\infty\!dt \int_{u_2}^\infty\!du_2\, \Big(-\frac{P_{u_1}}{\sigma}\Big)\Big|_{u_2=u_3=u}\,.
\end{align}
The function $P_\beta$, for arbitrary arguments, is presented in Appendix~\ref{App:Pbeta}. 
It is a combination of logarithms and simple rational functions;
thus, the integral in $u_2$ can readily be performed, and the result for $w=u$ is 
\begin{widetext}
\begin{align}
\GiB(t,u) =&\ -\int_t^\infty dt\,\frac{1}{4\,u^2}
 \left[\frac{g_1(t,u)}{t-2\,u}-\frac{g_1(t,u)}{t}+\frac{g_2(t,u)}{t}-\frac{g_2(t,u)}{t+2\,u}\right],\label{Eq:G1B}
\end{align}
where
\begin{align}
g_1(t,u)&=\frac{\pi ^2}{12}-\frac{1}{2}\ln^2\left(\frac{2\,u}{t+2\,u}\right)
 -\text{Li}_2\left(\frac{t}{t+2\,u}\right)+\text{Li}_2\left(\frac{t-2\,u}{t+2\,u}\right), \label{Eq:g1}\\
g_2(t,u)&=\frac{\pi ^2}{12}+\frac{1}{2}\ln^2\left(\frac{2\,u}{t+2\,u}\right)
 -2\,\text{Li}_2\left(\frac{t}{t+2\,u}\right)+\text{Li}_2\left(\frac{t-2\,u}{t+2\,u}\right).\label{Eq:g2}
\end{align}
Repeating the above derivation but with $w\neq u$, we obtain
\begin{align}
\GiB(t,u,w) &= -\int_t^\infty dt\,  \frac{\partial  \GiB(t,u,w)}{\partial t}\label{Eq:Gt1B}\\
\frac{\partial\GiB(t,u,w)}{\partial t} 
&=\frac{1}{4\,u\,w}\,\bigg[\frac{g_1(t,u,w)}{t-u-w}-\frac{g_1(t,u,w)}{t+u-w} 
                          +\frac{g_2(t,u,w)}{t-u+w}-\frac{g_2(t,u,w)}{t+u+w}\bigg]\label{Eq:tt5}
\end{align}
where
\begin{align}
g_1(t,u,w)&=\text{Li}_2\left(\frac{t-u-w}{t+u+w}\right)
 -\text{Li}_2\left(\frac{t+u-w}{t+u+w}\right)
 -\text{Li}_2\left(-\frac{u}{w}\right)-\frac{1}{2}\ln^2\left(\frac{2w}{t+u+w}\right), \label{Eq:g1tuw}\\
g_2(t,u,w)&=\text{Li}_2\left(\frac{t-u-w}{t+u+w}\right)-2\,\text{Li}_2\left(\frac{t-u+w}{t+u+w}\right)
 +\text{Li}_2\left(-\frac{u}{w}\right)+\frac{1}{2}\ln^2\left(\frac{2w}{t+u+w}\right)
 +\frac{\pi ^2}{6}\,. \label{Eq:g2tuw}
\end{align}
For the recurrence relations discussed in the following section, we will also need
derivatives of the master integral with respect to $u$ and $w$
\begin{align}
\frac{\partial\GiB(t,u,w)}{\partial u} 
&=-\frac{\GiB(t,u,w)}{u}-\frac{1}{4\,u\,w}\,\bigg[ 
 \frac{g_1(t,u,w)}{t-u-w}+\frac{g_1(t,u,w)}{t+u-w} 
+\frac{g_2(t,u,w)}{t-u+w}+\frac{g_2(t,u,w)}{t+u+w} \bigg], \label{Eg:dG1Bdu} \\
\frac{\partial\GiB(t,u,w)}{\partial w}
&=-\frac{\GiB(t,u,w)}{w}-\frac{1}{4\,u\,w}\,\bigg[ 
 \frac{g_1(t,u,w)}{t-u-w}-\frac{g_1(t,u,w)}{t+u-w} 
-\frac{g_2(t,u,w)}{t-u+w}+\frac{g_2(t,u,w)}{t+u+w} \bigg].\label{Eg:dG1Bdw}
\end{align}
These derivatives can be obtained from Eqs.~(\ref{Eq:Gt1B}) and~(\ref{Eq:tt5}) as follows
\begin{align}
\frac{\partial\GiB(t,u,w)}{\partial w}
&=\frac{\partial}{\partial w}\left(-\int_t^\infty\!dt\,\frac{\partial\GiB(t,u,w)}{\partial t}\right)  \\ 
&=- \int_t^\infty\!dt\, \frac{\partial}{\partial w}\frac{1}{4\,u\,w}\,\bigg[
 \frac{g_1(t,u,w)}{t-u-w}-\frac{g_1(t,u,w)}{t+u-w} + \frac{g_2(t,u,w)}{t-u+w}
 -\frac{g_2(t,u,w)}{t+u+w} \bigg] \nn \\ 
&= -\frac{\GiB(t,u,w)}{w}  - \frac{1}{4\,u\,w}\, \int_t^\infty\!dt\,
 \frac{\partial}{\partial w}\bigg[ \frac{g_1(t,u,w)}{t-u-w}-\frac{g_1(t,u,w)}{t+u-w} 
 + \frac{g_2(t,u,w)}{t-u+w} - \frac{g_2(t,u,w)}{t+u+w} \bigg] \nn \\ 
&= -\frac{\GiB(t,u,w)}{w} - \frac{1}{4\,u\,w}\, \int_t^\infty\!dt\bigg\{
 \frac{\partial}{\partial t}\bigg[ -\frac{g_1(t,u,w)}{t-u-w}+\frac{g_1(t,u,w)}{t+u-w} 
 + \frac{g_2(t,u,w)}{t-u+w} - \frac{g_2(t,u,w)}{t+u+w} \bigg] +X\bigg\} \nn
\end{align}
\end{widetext}
where
\begin{align}
&X=\bigg(\frac{1}{t-u-w} - \frac{1}{t+u-w}\bigg)\,\bigg( \frac{\partial}{\partial w} 
 +\frac{\partial}{\partial t}\bigg)\,g_1(t,u,w)\nn\\
&+ \bigg(\frac{1}{t-u+w}
 - \frac{1}{t+u+w}\bigg)\,\bigg( \frac{\partial}{\partial w} 
 - \frac{\partial}{\partial t}\bigg)\,g_2(t,u,w) \equiv 0
\end{align}
Hence, Eq.~(\ref{Eg:dG1Bdw}) is proved. 

The direct integral representation of $\GiB(t,u,w)$, namely
\begin{align}
\GiB(t,u,w) &= \int_{0}^\infty\!dk\, g(t,u+k,u,w)\,, \label{Eq:dkg}
\end{align}
where
\begin{align}
&g(t,u_2,u_3,w)= \frac{1}{2\,w(u_2-u_3) (u_2+u_3)}\nn\\
&\quad\times\biggl[
- \Li\biggl(\frac{t - u_3 - w}{t + u_2 + w}\biggr) + \Li\biggl( \frac{t - u_3 + w}{t + u_2 + w}\biggr) \nn\\
&\hspace{6ex} + \Li\biggl(\frac{t+u_2-w}{t+u_2+w}\biggr) + \Li\biggl(\frac{t-u_2-w}{t+u_3+w}\biggr) \nn\\
&\hspace{6ex} - \Li\biggl(\frac{t-u_2+w}{t+u_3+w}\biggr) - \Li\biggl(\frac{t+u_3-w}{t+u_3+w}\biggr)\biggr],
 \label{Eq:gtu2u3w}
\end{align}
will also be needed for deriving the recurrences. The latter formula was obtained using 
\begin{align}
g(t,u_2,u_3,w)&=-\int_t^\infty\!dt\,\frac{\partial g}{\partial t}(t,u_2,u_3,w)
\end{align}
with the integrand taken from Eq.~(\ref{Eq:dgdt}).

\subsection{Recurrences}
\label{Sec:Rec1B}

In this section we are going to derive formulas for $\GiB(t,u;\{n_i\})$ in Eq.~(\ref{EtildeG}) for arbitrary $n_i$, and
the subsequent steps of this derivation are as follows.

\subsubsection{Recurrence in \texorpdfstring{$n_3$}{n3}}
In the first step, we obtain formulas for the standard integral $g(t,w_1,x,u_2,u_3,w; n_3)$ at $w_1=x=0$.
For this purpose, we employ the PDE~(\ref{PDE}) with $\beta=x$.
For $w_1=0$, we take 
\begin{align}
\sigma = 16\,t^2\,u\,w\,x\,y - 16\,(-w\,x + u\,y) (u\,x - w\,y) (u\,w - x\,y)
\end{align}
and $P_x=P_{w_2}-P_{w_3}$ (see Appendix~\ref{App:Pbeta}). Next, we differentiate 
this PDE $n_3$ times with respect to $x$, set $x=0$, and extract the highest-order derivative 
\begin{align}
g(n_3)\equiv g(t,w_1,x,u_2,u_3,w; n_3)\big|_{x=0} = (-1)^{n_3}\dfrac{\partial^{n_3} g}{\partial x^{n_3}}\biggr|_{x=0}.
\end{align} 
The obtained recursive formula enables the $n_3$-th derivative $g(n_3)$ to be evaluated 
from lower-order derivatives of $g$ and the derivative of the inhomogeneous term 
$P_x(n_3) \equiv (-1)^{n_3}\dfrac{\partial^{n_3} P_x}{\partial x^{n_3}}\biggr|_{x=0}$
\begin{align}\label{Eq:gn3}
g(n_3)&=-\frac{(2\,n_3-1)\left(t^2-u^2-w^2-y^2\right)}{2\,u\,w\,y}\,g(n_3-1)\nn\\
&\quad-\frac{(n_3-1)^2\left(u^2 w^2+u^2 y^2+w^2 y^2\right)}{u^2\,w^2\,y^2}\, g(n_3-2\nn\\
&\quad-\frac{(n_3-2)(2\,n_3-3)(n_3-1)}{2\,u\,w\,y}\,g(n_3-3)\nn\\
&\quad+\frac{P_x(n_3-1)}{16\,u^2\,w^2\,y^2}\,.
\end{align}
The $g(n_3)$ obtained from this relation is the starting point for the next recurrence.

\subsubsection{Recurrence in \texorpdfstring{$n_1$}{n1}}
In the second step, we will obtain formulas for  $g(t,w_1,x,u_2,u_3,w; n_1,n_3)$ at $w_1=x=0$.
In a similar way as above, we begin the PDE~(\ref{PDE}) with $\beta=w_1$,
$\sigma$ from Eq.~(\ref{Eq:sigma}),
and
$P_{w_1}$ taken from Appendix~\ref{App:Pbeta}.
We differentiate this equation $n_1$ times with respect to $w_1$ and set $w_1=0$. 
Then, we differentiate the obtained relationship again $n_3$ times with respect to $x$ and set $x=0$.
These operations yield
\begin{align}
g(n_1,n_3) \equiv (-1)^{n_1+n_3}\dfrac{\partial^{n_3}}{\partial x^{n_3}}\dfrac{\partial^{n_1}}{\partial w_1^{n_1}}\,g\biggr|_\text{\parbox[b]{4em}{\setlength{\baselineskip}{1pt}$w_1=0\\x=0$}} 
\end{align}
with the following recursion relations:
\begin{widetext}
\begin{align}
&g(n_1,n_3)\\
&=\frac{c_1\,(n_1-1)^2\,(n_3-1)\,n_3\,g(n_1-2,n_3-2)}{8\,u^2\,w^2\,y^2}
 -\frac{c_2\,(n_1-1)^2\,g(n_1-2,n_3)}{16\,u^2\,w^2\,y^2}
 -\frac{c_3\,(n_3-1)\,n_3\,g(n_1,n_3-2)}{u^2\,w^2\,y^2}\nonumber\\
&\quad
 -\frac{c_4\,n_3\,g(n_1,n_3-1)}{u\,w\,y}
 -\frac{(n_1-3)\,(n_1-2)^2\,(n_1-1)\,t^2\,g(n_1-4,n_3)}{16\,u^2\,w^2\,y^2}
 -\frac{(n_3-2)\,(n_3-1)\,n_3\,g(n_1,n_3-3)}{u\,w\,y}\nonumber\\
&\quad-\frac{(n_1-1)^2\,(n_3-3)\,(n_3-2)\,(n_3-1)\,n_3\,g(n_1-2,n_3-4)}{16\,u^2\,w^2\,y^2}+\frac{(n_1-1)^2\,n_3\,g(n_1-2,n_3-1)}{2\,u\,w\,y}
 +\frac{P_{w_1}(n_1-1,n_3)}{16\,u^2\,w^2\,y^2}\,,\nn
\end{align}
where 
\begin{align}
c_1&=t^2 + u^2 + w^2 + y^2,&
c_2&=t^4 - 2\,t^2\,u^2 + u^4 - 2\,t^2\,w^2 - 2\,u^2\,w^2 + w^4 - 2\,t^2\,y^2 - 2\,u^2\,y^2 - 2\,w^2\,y^2 + y^4,\nn\\
c_3&=u^2\,w^2 + u^2\,y^2 + w^2\,y^2,&
c_4&=-t^2 + u^2 + w^2 + y^2.
\end{align}
\end{widetext}
and where we have defined $g(0,n_3)\equiv g(n_3)$ of Eq.~(\ref{Eq:gn3}) and $P_{w_1}(n_1,n_3)\equiv (-1)^{n_1+n_3}\dfrac{\partial^{n_3}}{\partial x^{n_3}}\dfrac{\partial^{n_1} }{\partial w_1^{n_1}}P_{w_1}\biggr|_\text{\parbox[b]{4em}{\setlength{\baselineskip}{1pt}$w_1=0\\x=0$}}$.

\subsubsection{Integration with respect to \texorpdfstring{$u_2$}{u2}}
\label{Subsec:intk}
In the third step, we perform analytic integration of $g(t,u_2,u_3,w;n_1,n_3)$
with respect to $u_2$ in order to obtain a function with an additional power of $1/r_{1B}$
\begin{align}
&\GiB(t,u,w;0,n_1,0,n_3,0,0)\nn\\
&\qquad=\int_{0}^\infty\!dk\,g(t,u+k,u,w;n_1,n_3)\biggr|_{w_1=x=0} .
\end{align}
The integrand combines dilogarithmic $(\Li)$, logarithmic, and rational functions of $t,u,w$, and $k$. 
The $\Li$ functions always appear in the same combination as in Eq.~(\ref{Eq:gtu2u3w}), namely, 
they form $g(t,u+k,u,w)$, and we use this equation to express the integral in terms of $\GiB(t,u,w)$ 
according to Eq.~(\ref{Eq:dkg}). Further on, the integration of logarithmic functions gives 
dilogarithms $\Li$ in such a combination, which can always be expressed in terms of $g_1$ and $g_2$ 
functions defined in Eqs.~(\ref{Eq:g1tuw}) and (\ref{Eq:g2tuw}). What remains after the integration 
are the logarithmic and rational functions of $t$, $u$, and $w$. 

\subsubsection{Recurrence in \texorpdfstring{$n_4$}{n4} and \texorpdfstring{$n_5$}{n5}}
In the fourth step, we derive $\GiB(t,u;0,n_1,0,n_3,n_4,n_5)$ by taking derivatives with respect to $u$ and $w$ of $\GiB(t,u,w;0,n_1,0,n_3,0,0)$
\begin{align}
&\GiB(t,u;0,n_1,0,n_3,n_4,n_5)\\
&=
\!\left(-\frac{\partial}{\partial w}\right)^{n_5}_{w=u}
\!\left(-\frac{\partial}{\partial u}\right)^{n_4}
\GiB(t,u,w;0,n_1,0,n_3,0,0),\nn
\end{align}
and in this operation we make use of Eqs.~(\ref{Eg:dG1Bdu}) and (\ref{Eg:dG1Bdw}).

\subsubsection{Recurrence in \texorpdfstring{$n_2$}{n2}}

Recurrence in the $n_2$ exponent can be found from an algebraic relation between variables
\begin{equation}
\frac{\eta_1^{n_2}}{\riB}
 =\eta_1^{n_2-1}\left(\frac{\zeta_1}{\riB}-2\right) ,
\end{equation}
which leads directly to the following formula:
\begin{align}
&\GiB(t,u;0,n_1,n_2,n_3,n_4,n_5)\nn\\
&\qquad=\GiB(t,u;0,n_1,n_2-1,n_3,n_4+1,n_5)\nn\\
&\qquad\quad-2\,G(t,u;0,n_1,n_2-1,n_3,n_4,n_5).
\end{align}
As we can see, apart from $\GiB$ integrals of lower order in $n_2$, it involves also
standard integrals $G$ of Eq.~(\ref{Eq:Gtun}).

\subsubsection{Recurrence in \texorpdfstring{$n_0$}{n0}}

The last exponent for which we need to find a recurrence is $n_0$, which  
relates to the internuclear variable $R$ and the non-linear parameter $t$. 
Because of the presence of the parameter $t$ in the naJC 
basis function, Eq.~(\ref{Eq:naJC}), the $n_0=0$ integrals
have an explicit dependence on $t$, e.g.,
\begin{align}
&\GiB(t,u;0,0,0,0,0,1)\nn\\
&\quad=\frac{1}{2\,t\,u}\left[2\,t\,\GiB(t,u)+
   \frac{g_1(t,u)}{t-2 u}-\frac{g_2(t,u)}{t+2 u}\right],\\
&\GiB(t,u;0,1,0,0,0,0)\nn\\
&\quad=\frac{1}{2\,t\,u\,(t+2 u)}\ln \left(\dfrac{2 u}{t+2 u}\right),\\
&\GiB(t,u;0,1,1,0,0,0)\nn\\
&\quad=\frac{1}{t\,(t+2 u)^2}\left[\frac{1}{2\,u}+\frac{1}{t}\ln \left(\frac{2 u}{t+2 u}\right)\right].
\end{align}
Therefore, it is sufficient to perform a direct differentiation of pertinent 
$\GiB$ integrals with respect to this variable
\begin{align}
\GiB(t,u;\{n_i\})&=
\left(-\frac{\partial}{\partial t}\right)^{n_0}\!\GiB(t,u;0,n_1,n_2,n_3,n_4,n_5)
\end{align}
using Eq.~(\ref{Eq:G1B}) to obtain formulas for an arbitrary $n_0$. 
This concludes the derivation of explicit formulas for an arbitrary $\GiB$ integral.
A few examples of moderate-size formulas are given below:
\begin{align}
&\GiB(t,u;1,0,0,0,0,1)\nn\\
&=-\frac{t^2 - 4\,t\,u + 2\,u^2}{2\,t^2\,u^2\,(t-2\,u)^2}\,g_1(t,u)
 -\frac{t^2 + 4\,t\,u + 2\,u^2}{2\,t^2\,u^2\,(t+2\,u)^2}\,g_2(t,u) \nn\\
&\quad+\frac{8\,u}{t\,(t - 2 u)^2 (t + 2 u)^2}\,\ln{2}\nn\\
&\quad -\frac{t^3 - 3\,t^2\,u - 12\,u^3}{t^2\,u\,(t - 2\,u)^2 (t + 2\,u)^2}\,\ln\frac{2\,u}{t+2\,u}\\[0.5em]
&\GiB(t,u;1,1,0,0,0,0)\nn\\
&=\frac{1}{2\,t\,u\,(t+2\,u)^2}\left[
1+\frac{2\,(t+u)}{t}\ln{\frac{2\,u}{t+2\,u}}\right],\\[0.5em]
&\GiB(t,u;1,1,1,0,0,0)\nn\\
&=\frac{1}{t^2\,(t+2\,u)^3}\left[
\frac{3\,t+4\,u}{2\,u}+\frac{4\,(t+u)}{t}\ln{\frac{2\,u}{t+2\,u}}
\right].
\end{align}

\section{Numerical results}

\begin{table}[!htb]
\caption{Numerical values of the master integrals,
defined in Eqs.~(\ref{EhatGm}),~(\ref{EddotGm}), and~(\ref{EtildeGm}), 
evaluated for $t=38.38$ and $u=1.956$, are shown with a precision 
of 32 significant digits.}
\label{Tab:mi}
\begin{center}
\begin{tabular}{l@{\quad}l}
\hline\hline
Master integral & Value \\
\hline
$\GAB(t,u)$& 0.007\,321\,991\,591\,821\,939\,899\,610\,091\,538\,52 \\
$\Gij(t,u)$& 0.002\,007\,747\,417\,108\,337\,201\,534\,734\,124\,51 \\
$\GiB(t,u)$& 0.002\,832\,386\,807\,422\,208\,953\,576\,913\,409\,21 \\
\hline\hline
\end{tabular}
\end{center}
\end{table}

In this section, we present a small selection of numerical results that can be helpful
in reproduction and numerical implementation of the equations derived
in previous sections. Among many formulas employed to produce
the full set of the relativistic integrals, those for the master
integrals seem to be the most important. Because they are the seeds 
of all recurrences, their numerical values must be known to a sufficiently
high precision. Every step of the recurrence in $n_i$ may introduce a small 
round-off error, which when accumulated would deteriorate the precision 
of the highest order terms. Because the target precision imposed on all the integrals
is about 64 digits, the master integrals must be evaluated to a significantly 
higher accuracy. This goal has been achieved using MPFR libraries \cite{MPFR:07}
coupled with a MPFUN library \cite{MPFUN:20} and linked to a source code in Fortran 95.
Numerical values of the master integrals representing three different classes
of relativistic integrals are listed in Table~\ref{Tab:mi}.

The total energy of a rovibrational level of a light molecule described by the vibrational $v$ 
and rotational $J$ quantum numbers is represented as a series in powers of the fine-structure constant $\alpha$
\begin{align}\label{Eq:alpha}
E^{(v,J)}&=\alpha^2\,E_\mathrm{nr}^{(v,J)}+\alpha^4\,E_\mathrm{rel}^{(v,J)}+\alpha^5\,E_\mathrm{qed}^{(v,J)}
 +\dots\,.
\end{align}
Our ultimate purpose, for which the integrals described above are indispensable, is
an accurate prediction of the relativistic correction $E_\mathrm{rel}^{(v,J)}$ for rovibrational states 
of H$_2$ and its isotopologues. This correction is evaluated as an expectation value
$\langle \Psi | H_\mathrm{BP}| \Psi \rangle$ of the mass-dependent Breit-Pauli 
Hamiltonian (in atomic units, $m=1$)
\begin{widetext}
\begin{align}
H_\mathrm{BP}&=-\frac{p_1^4}{8\,m^3} - \frac{p_2^4}{8\,m^3} - \frac{p_A^4}{8\,m_A^3} - \frac{p_B^4}{8\,m_B^3} 
 +\frac{\pi}{m^2}\,\delta^{(3)}(r_{12})
 +\frac{\pi}{2}\left(\frac{1}{m^2}+\frac{\delta_{I_A}}{m_A^2}\right)\left(\delta^{(3)}(r_{1A})+\delta^{(3)}(r_{2A})\right)\nn\\
&\quad +\frac{\pi}{2}\left(\frac{1}{m^2}+\frac{\delta_{I_B}}{m_B^2}\right)\left(\delta^{(3)}(r_{1B})+\delta^{(3)}(r_{2B})\right)
 -\frac{1}{2\,m^2}\,p_1^i\,\biggl(\frac{\delta^{ij}}{r_{12}} +\frac{r_{12}^i\,r_{12}^j}{r_{12}^3}\biggr)\,p_2^j 
 -\frac{1}{2\,m_A\,m_B}\,p_A^i\,\biggl(\frac{\delta^{ij}}{r_{AB}} 
 + \frac{r_{AB}^i\,r_{AB}^j}{r_{AB}^3}\biggr)\,p_B^j
 \nn\\ &\quad
 +\frac{1}{2\,m\,m_A}\,p_1^i\,\biggl(\frac{\delta^{ij}}{r_{1A}} 
 + \frac{r_{1A}^i\,r_{1A}^j}{r_{1A}^3}\biggr)\,p_A^j
 +\frac{1}{2\,m\,m_B}\,p_1^i\,\biggl(\frac{\delta^{ij}}{r_{1B}} 
 + \frac{r_{1B}^i\,r_{1B}^j}{r_{1B}^3}\biggr)\,p_B^j\nn\\
&\quad+\frac{1}{2\,m\,m_A}\,p_2^i\,\biggl(\frac{\delta^{ij}}{r_{2A}} 
 + \frac{r_{2A}^i\,r_{2A}^j}{r_{2A}^3}\biggr)\,p_A^j
 +\frac{1}{2\,m\,m_B}\,p_2^i\,\biggl(\frac{\delta^{ij}}{r_{2B}} 
 + \frac{r_{2B}^i\,r_{2B}^j}{r_{2B}^3}\biggr)\,p_B^j  \label{Eq:BP}
\end{align}
\end{widetext}
with the wavefunction $\Psi$ expanded in the basis of the nonadiabatic James-Coolidge (naJC) functions,
Eq.~(\ref{Eq:naJC}). In the above equation, subscripts $A$ and $B$, accompanying symbols of mass ($m$), 
momentum ($p$), and coordinate ($r$), concern nuclei, while $1$ and $2$ refer to electrons.
The nuclear-spin factor $\delta_I$, present in Dirac delta terms, depends on the nucleus' spin $I$: 
$\delta_I=1$ for $I=1/2$ and $\delta_I=0$ otherwise. All the electron spin-dependent terms are omitted 
as they vanish for the ground electronic state of $^1\Sigma^+_g$ symmetry,
while nuclear-spin dependent terms are also omitted because we do not consider the fine and hyperfine structure.
Due to its negligible magnitude, we have also omitted the nucleus-nucleus Dirac delta term.

Table~\ref{Tab:conv} contains preliminary numerical results of the relativistic correction 
obtained for the three lowest rovibrational levels of H$_2$. $E_\mathrm{rel}^{(v,J)}$ was evaluated with 
a sequence of wavefunctions of growing quality, which enables estimation of its numerical accuracy.
The size of the wavefunction expansions was determined by the shell parameter $\Omega$
limiting from above the sum of the exponents $n_1+n_2+n_3+n_4+n_5$ of the naJC basis functions~(\ref{Eq:naJC})
included.
The extrapolation to the infinite basis size was performed at the level of individual
operators present in the Hamiltonian~(\ref{Eq:BP}).
The relativistic integrals were evaluated for integer exponents fulfilling the following conditions
$n_1+n_2+n_3+n_4+n_5\leq 35$ and $n_0\le 85$, which enables application of wavefunctions
with the shell $\Omega$ up to 14 for $J=0$ and up to 12 for $J>0$.

For the rotationless level $(J=0)$, analogous results are available in the literature. 
In 2018, Wang and Yan \cite{Wang:18a} reported $E_\mathrm{rel}^{(0,0)}=-0.204\,544(5)$ a.u. 
in agreement with our results, whereas Puchalski {\em et al.} \cite{PSKP:18} obtained 
$E_\mathrm{rel}^{(0,0)}=-0.204\,547\,56(4)$ a.u.,
which is off by $4\,\sigma$ from the new result. Reinvestigating the convergence
of the latter correction revealed that the error bar estimation was too optimistic.
Calculations performed by Stanke and Adamowicz \cite{Stanke:13} in 2013 yielded 
$E_\mathrm{rel}^{(0,0)}=-0.201\,3$ a.u. The uncertainty of this number is unknown.
Assuming that all the digits quoted are significant, we note a considerable disagreement 
with all the other values.

\begin{widetext}
\begin{table*}[!htb]
\caption{Convergence of the relativistic correction $E_\mathrm{rel}^{(v,J)}$ (in a.u.) calculated 
using the nonadiabatic James-Coolidge (naJC) wavefunction for the $(v,J)$ rovibrational level of H$_2$.
$K$ is the size of the naJC basis set employed, governed by $\Omega$ -- the largest shell enabled. Calculations were performed using the nuclear mass $M/m=1836.152\,673\,43(11)$\ \cite{CODATA:18}.}
\label{Tab:conv}
\begin{center}
\begin{tabular*}{\textwidth}{rr@{\extracolsep{\fill}}rx{2.12}rx{2.12}rx{2.12}}
\hline\hline
&$\Omega$ & $K\quad$ & \cent{E_\mathrm{rel}^{(0,0)}} & $K\quad$ & \cent{E_\mathrm{rel}^{(0,1)}} & $K\quad$ & \cent{E_\mathrm{rel}^{(0,2)}} \\
\hline
&  9 &  28\,756 & -0.204\,547\,752\,0   &  49\,042 & -0.204\,326\,998\,3   &  49\,042 & -0.203\,890\,204\,8 \\
& 10 &  42\,588 & -0.204\,547\,538\,4   &  73\,164 & -0.204\,326\,718\,0   &  73\,164 & -0.203\,889\,953\,8 \\
& 11 &  61\,152 & -0.204\,547\,467\,0   & 105\,840 & -0.204\,326\,616\,4   & 105\,840 & -0.203\,889\,881\,7 \\
& 12 &  85\,904 & -0.204\,547\,434\,3   & 149\,408 & -0.204\,326\,587\,7   & 149\,408 & -0.203\,889\,846\,0\\
& 13 & 117\,936 & -0.204\,547\,423\,1   &  $-\ $   &\multicolumn{1}{c}{$-$}& $-\ $    &\multicolumn{1}{c}{$-$}\\
& 14 & 159\,120 & -0.204\,547\,417\,6   &  $-\ $   &\multicolumn{1}{c}{$-$}& $-\ $    &\multicolumn{1}{c}{$-$}\\
& &$\infty\quad$& -0.204\,547\,412(5)   &$\infty\ $& -0.204\,326\,56(3) &$\infty\ $   & -0.203\,889\,81(3)\\
\hline\hline
\end{tabular*}
\end{center}
\end{table*}
\end{widetext}

For the rotationally excited levels there are no analogous data available in the literature. 
A comparison can be made to the relativistic correction obtained within the adiabatic approximation, 
e.g., within the nonadiabatic perturbation theory (NAPT) implemented in the publicly available H2Spectre program 
\cite{Komasa:19,H2Spectre}. For $J=1$, NAPT yields $E_\mathrm{rel}^{(0,1)}=-0.204\,326\,8(2)$ a.u.,
which agrees to within $1.2\,\sigma$ with the direct nonadiabatic (DNA) result of Table~\ref{Tab:conv}. 
The uncertainty of the NAPT result is due to neglected higher order finite nuclear mass effects;
the comparison with the DNA value validates the method of uncertainty estimation.
For $J=2$, NAPT gives $E_\mathrm{rel}^{(0,2)}=-0.203\,889\,6(2)$ a.u., which is in agreement
with the DNA result.

\section{Conclusions}

The naJC wavefunction, together with the nuclear-mass dependent Breit-Pauli Hamiltonian 
of Eq.~(\ref{Eq:BP}), fully take into account nonadiabatic effects (nuclear recoil) in 
the relativistic correction. However, the expectation values of the operators present 
in this Hamiltonian evaluated in the naJC basis require access to new, previously 
unknown classes of integrals. The mathematical techniques reported in this paper enabled 
the evaluation of such extended integrals, allowing an unprecedented relative accuracy 
of $3\cdot 10^{-8}$ for the relativistic correction of the ground state of H$_2$. 
Regarding the dissociation energy of a rovibrational level, this corresponds to 
an absolute accuracy of $6\cdot 10^{-8}\,\icm$ ($\sim\!2$ kHz).
Previous calculations, apart from the rotationless cases mentioned above, were performed 
in the framework of the adiabatic approximation using the second-order nonadiabatic perturbation 
theory (NAPT) with the inclusion of the relativistic terms proportional to 
the electron-to-nucleus mass ratio. The two-orders-of-magnitude more accurate new DNA method 
removes the uncertainty caused by unknown higher-order terms of the NAPT expansion and enables 
the error estimation of the NAPT computation to be verified. 

One of the essential features opened up by the extended classes of integrals 
is the possibility of accurately determining the relativistic correction for higher rotational levels.
As in the case of the nonrelativistic energy \cite{PK:16,PK:18a,PK:18b,PK:19,PK:22}, 
the accuracy now achieved allows the error from the relativistic correction 
to be neglected in the total error budget.
From now on, the missing recoil contribution to the QED correction and the unknown higher-order 
in $\alpha$ corrections will be the only factors that determine the overall energy uncertainty.

Apart from the relativistic correction itself,
the new classes of integrals will also enable an extension of the field of application
of the naJC wavefunction to the evaluation of the operators present in the 
QED term of the expansion~(\ref{Eq:alpha}) as well as to various electric and magnetic
properties of the hydrogen molecule.

\section*{Acknowledgment}
\noindent This research was supported by the National Science Center (Poland) Grant No. 2021/41/B/ST4/00089. 
A~computer grant from the Poznań Supercomputing and Networking Center was used to carry out the numerical calculations.

\bibliographystyle{apsrev}  

\begin{thebibliography}{41}
\expandafter\ifx\csname natexlab\endcsname\relax\def\natexlab#1{#1}\fi
\expandafter\ifx\csname bibnamefont\endcsname\relax
  \def\bibnamefont#1{#1}\fi
\expandafter\ifx\csname bibfnamefont\endcsname\relax
  \def\bibfnamefont#1{#1}\fi
\expandafter\ifx\csname citenamefont\endcsname\relax
  \def\citenamefont#1{#1}\fi
\expandafter\ifx\csname url\endcsname\relax
  \def\url#1{\texttt{#1}}\fi
\expandafter\ifx\csname urlprefix\endcsname\relax\def\urlprefix{URL }\fi
\providecommand{\bibinfo}[2]{#2}
\providecommand{\eprint}[2][]{\url{#2}}

\bibitem[{\citenamefont{Fraser et~al.}(2002)\citenamefont{Fraser, McCoustra,
  and Williams}}]{Fraser:02}
\bibinfo{author}{\bibfnamefont{H.~J.} \bibnamefont{Fraser}},
  \bibinfo{author}{\bibfnamefont{M.~R.~S.} \bibnamefont{McCoustra}},
  \bibnamefont{and} \bibinfo{author}{\bibfnamefont{D.~A.}
  \bibnamefont{Williams}}, \bibinfo{journal}{Astronomy \& Geophysics}
  \textbf{\bibinfo{volume}{43}}, \bibinfo{pages}{2.10} (\bibinfo{year}{2002}),
  ISSN \bibinfo{issn}{1366-8781}.

\bibitem[{\citenamefont{Margolis and Hunt}(1973)}]{Margolis:73}
\bibinfo{author}{\bibfnamefont{J.}~\bibnamefont{Margolis}} \bibnamefont{and}
  \bibinfo{author}{\bibfnamefont{G.}~\bibnamefont{Hunt}},
  \bibinfo{journal}{Icarus} \textbf{\bibinfo{volume}{18}}, \bibinfo{pages}{593}
  (\bibinfo{year}{1973}), ISSN \bibinfo{issn}{0019-1035},
  \urlprefix\url{https://www.sciencedirect.com/science/article/pii/0019103573900614}.

\bibitem[{\citenamefont{{Roueff, E.} et~al.}(2019)\citenamefont{{Roueff, E.},
  {Abgrall, H.}, {Czachorowski, P.}, {Pachucki, K.}, {Puchalski, M.}, and
  {Komasa, J.}}}]{Roueff:19}
\bibinfo{author}{\bibnamefont{{Roueff, E.}}},
  \bibinfo{author}{\bibnamefont{{Abgrall, H.}}},
  \bibinfo{author}{\bibnamefont{{Czachorowski, P.}}},
  \bibinfo{author}{\bibnamefont{{Pachucki, K.}}},
  \bibinfo{author}{\bibnamefont{{Puchalski, M.}}}, \bibnamefont{and}
  \bibinfo{author}{\bibnamefont{{Komasa, J.}}}, \bibinfo{journal}{Astron.
  Astrophys.} \textbf{\bibinfo{volume}{630}}, \bibinfo{pages}{A58}
  (\bibinfo{year}{2019}).

\bibitem[{\citenamefont{Tchernyshyov}(2022)}]{Tchernyshyov:22}
\bibinfo{author}{\bibfnamefont{K.}~\bibnamefont{Tchernyshyov}},
  \bibinfo{journal}{Astrophys. J.} \textbf{\bibinfo{volume}{931}},
  \bibinfo{pages}{78} (\bibinfo{year}{2022}),
  \urlprefix\url{https://dx.doi.org/10.3847/1538-4357/ac68e0}.

\bibitem[{\citenamefont{Tan et~al.}(2022)\citenamefont{Tan, Skinner, Samuels,
  Hargreaves, Hashemi, and Gordon}}]{Tan:22}
\bibinfo{author}{\bibfnamefont{Y.}~\bibnamefont{Tan}},
  \bibinfo{author}{\bibfnamefont{F.~M.} \bibnamefont{Skinner}},
  \bibinfo{author}{\bibfnamefont{S.}~\bibnamefont{Samuels}},
  \bibinfo{author}{\bibfnamefont{R.~J.} \bibnamefont{Hargreaves}},
  \bibinfo{author}{\bibfnamefont{R.}~\bibnamefont{Hashemi}}, \bibnamefont{and}
  \bibinfo{author}{\bibfnamefont{I.~E.} \bibnamefont{Gordon}},
  \bibinfo{journal}{Astrophys. J. Supp. Ser.} \textbf{\bibinfo{volume}{262}},
  \bibinfo{pages}{40} (\bibinfo{year}{2022}).

\bibitem[{\citenamefont{Gordon et~al.}(2022)\citenamefont{Gordon, Rothman,
  Hargreaves, Hashemi, Karlovets, Skinner, Conway, Hill, Kochanov, Tan
  et~al.}}]{Gordon:22}
\bibinfo{author}{\bibfnamefont{I.}~\bibnamefont{Gordon}},
  \bibinfo{author}{\bibfnamefont{L.}~\bibnamefont{Rothman}},
  \bibinfo{author}{\bibfnamefont{R.}~\bibnamefont{Hargreaves}},
  \bibinfo{author}{\bibfnamefont{R.}~\bibnamefont{Hashemi}},
  \bibinfo{author}{\bibfnamefont{E.}~\bibnamefont{Karlovets}},
  \bibinfo{author}{\bibfnamefont{F.}~\bibnamefont{Skinner}},
  \bibinfo{author}{\bibfnamefont{E.}~\bibnamefont{Conway}},
  \bibinfo{author}{\bibfnamefont{C.}~\bibnamefont{Hill}},
  \bibinfo{author}{\bibfnamefont{R.}~\bibnamefont{Kochanov}},
  \bibinfo{author}{\bibfnamefont{Y.}~\bibnamefont{Tan}}, \bibnamefont{et~al.},
  \bibinfo{journal}{J. Quant. Spectrosc. Radiat. Transfer}
  \textbf{\bibinfo{volume}{277}}, \bibinfo{pages}{107949}
  (\bibinfo{year}{2022}), ISSN \bibinfo{issn}{0022-4073},
  \urlprefix\url{https://www.sciencedirect.com/science/article/pii/S0022407321004416}.

\bibitem[{\citenamefont{Sung et~al.}(2023)\citenamefont{Sung, Wishnow, Drouin,
  Manceron, Verseils, Benner, and Nixon}}]{Sung:23}
\bibinfo{author}{\bibfnamefont{K.}~\bibnamefont{Sung}},
  \bibinfo{author}{\bibfnamefont{E.~H.} \bibnamefont{Wishnow}},
  \bibinfo{author}{\bibfnamefont{B.~J.} \bibnamefont{Drouin}},
  \bibinfo{author}{\bibfnamefont{L.}~\bibnamefont{Manceron}},
  \bibinfo{author}{\bibfnamefont{M.}~\bibnamefont{Verseils}},
  \bibinfo{author}{\bibfnamefont{D.~C.} \bibnamefont{Benner}},
  \bibnamefont{and} \bibinfo{author}{\bibfnamefont{C.~A.} \bibnamefont{Nixon}},
  \bibinfo{journal}{J. Quant. Spectrosc. Radiat. Transfer}
  \textbf{\bibinfo{volume}{295}}, \bibinfo{pages}{108412}
  (\bibinfo{year}{2023}), ISSN \bibinfo{issn}{0022-4073}.

\bibitem[{\citenamefont{{Ochsenbein, F.}
  et~al.}(2000)\citenamefont{{Ochsenbein, F.}, {Bauer, P.}, and {Marcout,
  J.}}}]{Ochsenbein:00}
\bibinfo{author}{\bibnamefont{{Ochsenbein, F.}}},
  \bibinfo{author}{\bibnamefont{{Bauer, P.}}}, \bibnamefont{and}
  \bibinfo{author}{\bibnamefont{{Marcout, J.}}}, \bibinfo{journal}{Astron.
  Astrophys. Suppl. Ser.} \textbf{\bibinfo{volume}{143}}, \bibinfo{pages}{23}
  (\bibinfo{year}{2000}).

\bibitem[{\citenamefont{Tennyson et~al.}(2020)\citenamefont{Tennyson,
  Yurchenko, Al-Refaie, Clark, Chubb, Conway, Dewan, Gorman, Hill, Lynas-Gray
  et~al.}}]{Tennyson:20}
\bibinfo{author}{\bibfnamefont{J.}~\bibnamefont{Tennyson}},
  \bibinfo{author}{\bibfnamefont{S.~N.} \bibnamefont{Yurchenko}},
  \bibinfo{author}{\bibfnamefont{A.~F.} \bibnamefont{Al-Refaie}},
  \bibinfo{author}{\bibfnamefont{V.~H.} \bibnamefont{Clark}},
  \bibinfo{author}{\bibfnamefont{K.~L.} \bibnamefont{Chubb}},
  \bibinfo{author}{\bibfnamefont{E.~K.} \bibnamefont{Conway}},
  \bibinfo{author}{\bibfnamefont{A.}~\bibnamefont{Dewan}},
  \bibinfo{author}{\bibfnamefont{M.~N.} \bibnamefont{Gorman}},
  \bibinfo{author}{\bibfnamefont{C.}~\bibnamefont{Hill}},
  \bibinfo{author}{\bibfnamefont{A.}~\bibnamefont{Lynas-Gray}},
  \bibnamefont{et~al.}, \bibinfo{journal}{J. Quant. Spectrosc. Radiat.
  Transfer} \textbf{\bibinfo{volume}{255}}, \bibinfo{pages}{107228}
  (\bibinfo{year}{2020}), ISSN \bibinfo{issn}{0022-4073}.

\bibitem[{\citenamefont{Wcisło et~al.}(2016)\citenamefont{Wcisło, Gordon,
  Tran, Tan, Hu, Campargue, Kassi, Romanini, Hill, Kochanov
  et~al.}}]{Wcislo:16a}
\bibinfo{author}{\bibfnamefont{P.}~\bibnamefont{Wcisło}},
  \bibinfo{author}{\bibfnamefont{I.}~\bibnamefont{Gordon}},
  \bibinfo{author}{\bibfnamefont{H.}~\bibnamefont{Tran}},
  \bibinfo{author}{\bibfnamefont{Y.}~\bibnamefont{Tan}},
  \bibinfo{author}{\bibfnamefont{S.-M.} \bibnamefont{Hu}},
  \bibinfo{author}{\bibfnamefont{A.}~\bibnamefont{Campargue}},
  \bibinfo{author}{\bibfnamefont{S.}~\bibnamefont{Kassi}},
  \bibinfo{author}{\bibfnamefont{D.}~\bibnamefont{Romanini}},
  \bibinfo{author}{\bibfnamefont{C.}~\bibnamefont{Hill}},
  \bibinfo{author}{\bibfnamefont{R.}~\bibnamefont{Kochanov}},
  \bibnamefont{et~al.}, \bibinfo{journal}{J. Quant. Spectrosc. Radiat.
  Transfer} \textbf{\bibinfo{volume}{177}}, \bibinfo{pages}{75}
  (\bibinfo{year}{2016}), ISSN \bibinfo{issn}{0022-4073},
  \bibinfo{note}{xVIIIth Symposium on High Resolution Molecular Spectroscopy
  (HighRus-2015), Tomsk, Russia},
  \urlprefix\url{https://www.sciencedirect.com/science/article/pii/S0022407315302028}.

\bibitem[{\citenamefont{Puchalski et~al.}(2020)\citenamefont{Puchalski, Komasa,
  and Pachucki}}]{Puchalski:20}
\bibinfo{author}{\bibfnamefont{M.}~\bibnamefont{Puchalski}},
  \bibinfo{author}{\bibfnamefont{J.}~\bibnamefont{Komasa}}, \bibnamefont{and}
  \bibinfo{author}{\bibfnamefont{K.}~\bibnamefont{Pachucki}},
  \bibinfo{journal}{Phys. Rev. Lett.} \textbf{\bibinfo{volume}{125}},
  \bibinfo{pages}{253001} (\bibinfo{year}{2020}).

\bibitem[{\citenamefont{Puchalski et~al.}(2022)\citenamefont{Puchalski, Komasa,
  Spyszkiewicz, and Pachucki}}]{Puchalski:22}
\bibinfo{author}{\bibfnamefont{M.}~\bibnamefont{Puchalski}},
  \bibinfo{author}{\bibfnamefont{J.}~\bibnamefont{Komasa}},
  \bibinfo{author}{\bibfnamefont{A.}~\bibnamefont{Spyszkiewicz}},
  \bibnamefont{and} \bibinfo{author}{\bibfnamefont{K.}~\bibnamefont{Pachucki}},
  \bibinfo{journal}{Phys. Rev. A} \textbf{\bibinfo{volume}{105}},
  \bibinfo{pages}{042802} (\bibinfo{year}{2022}),
  \urlprefix\url{https://link.aps.org/doi/10.1103/PhysRevA.105.042802}.

\bibitem[{\citenamefont{Salumbides et~al.}(2013)\citenamefont{Salumbides,
  Koelemeij, Komasa, Pachucki, Eikema, and Ubachs}}]{Salumbides:13}
\bibinfo{author}{\bibfnamefont{E.~J.} \bibnamefont{Salumbides}},
  \bibinfo{author}{\bibfnamefont{J.~C.~J.} \bibnamefont{Koelemeij}},
  \bibinfo{author}{\bibfnamefont{J.}~\bibnamefont{Komasa}},
  \bibinfo{author}{\bibfnamefont{K.}~\bibnamefont{Pachucki}},
  \bibinfo{author}{\bibfnamefont{K.~S.~E.} \bibnamefont{Eikema}},
  \bibnamefont{and} \bibinfo{author}{\bibfnamefont{W.}~\bibnamefont{Ubachs}},
  \bibinfo{journal}{Phys. Rev. D} \textbf{\bibinfo{volume}{87}},
  \bibinfo{pages}{112008} (\bibinfo{year}{2013}).

\bibitem[{\citenamefont{Ubachs et~al.}(2016)\citenamefont{Ubachs, Koelemeij,
  Eikema, and Salumbides}}]{Ubachs:16}
\bibinfo{author}{\bibfnamefont{W.}~\bibnamefont{Ubachs}},
  \bibinfo{author}{\bibfnamefont{J.}~\bibnamefont{Koelemeij}},
  \bibinfo{author}{\bibfnamefont{K.}~\bibnamefont{Eikema}}, \bibnamefont{and}
  \bibinfo{author}{\bibfnamefont{E.}~\bibnamefont{Salumbides}},
  \bibinfo{journal}{J. Mol. Spectrosc.} \textbf{\bibinfo{volume}{320}},
  \bibinfo{pages}{1 } (\bibinfo{year}{2016}), ISSN \bibinfo{issn}{0022-2852}.

\bibitem[{\citenamefont{Hollik et~al.}(2020)\citenamefont{Hollik, Linster, and
  Tabet}}]{Hollik:20}
\bibinfo{author}{\bibfnamefont{W.}~\bibnamefont{Hollik}},
  \bibinfo{author}{\bibfnamefont{M.}~\bibnamefont{Linster}}, \bibnamefont{and}
  \bibinfo{author}{\bibfnamefont{M.}~\bibnamefont{Tabet}},
  \bibinfo{journal}{Eur. Phys. J. C} \textbf{\bibinfo{volume}{80}},
  \bibinfo{pages}{661} (\bibinfo{year}{2020}).

\bibitem[{\citenamefont{Altmann et~al.}(2018)\citenamefont{Altmann, Dreissen,
  Salumbides, Ubachs, and Eikema}}]{Altmann:18}
\bibinfo{author}{\bibfnamefont{R.~K.} \bibnamefont{Altmann}},
  \bibinfo{author}{\bibfnamefont{L.~S.} \bibnamefont{Dreissen}},
  \bibinfo{author}{\bibfnamefont{E.~J.} \bibnamefont{Salumbides}},
  \bibinfo{author}{\bibfnamefont{W.}~\bibnamefont{Ubachs}}, \bibnamefont{and}
  \bibinfo{author}{\bibfnamefont{K.~S.~E.} \bibnamefont{Eikema}},
  \bibinfo{journal}{Phys. Rev. Lett.} \textbf{\bibinfo{volume}{120}},
  \bibinfo{pages}{043204} (\bibinfo{year}{2018}).

\bibitem[{\citenamefont{Cheng et~al.}(2018)\citenamefont{Cheng, Hussels, Niu,
  Bethlem, Eikema, Salumbides, Ubachs, Beyer, H{\"o}lsch, Agner
  et~al.}}]{Cheng:18}
\bibinfo{author}{\bibfnamefont{C.}~\bibnamefont{Cheng}},
  \bibinfo{author}{\bibfnamefont{J.}~\bibnamefont{Hussels}},
  \bibinfo{author}{\bibfnamefont{M.}~\bibnamefont{Niu}},
  \bibinfo{author}{\bibfnamefont{H.~L.} \bibnamefont{Bethlem}},
  \bibinfo{author}{\bibfnamefont{K.~S.~E.} \bibnamefont{Eikema}},
  \bibinfo{author}{\bibfnamefont{E.~J.} \bibnamefont{Salumbides}},
  \bibinfo{author}{\bibfnamefont{W.}~\bibnamefont{Ubachs}},
  \bibinfo{author}{\bibfnamefont{M.}~\bibnamefont{Beyer}},
  \bibinfo{author}{\bibfnamefont{N.~J.} \bibnamefont{H{\"o}lsch}},
  \bibinfo{author}{\bibfnamefont{J.~A.} \bibnamefont{Agner}},
  \bibnamefont{et~al.}, \bibinfo{journal}{Phys. Rev. Lett.}
  \textbf{\bibinfo{volume}{121}}, \bibinfo{pages}{013001}
  (\bibinfo{year}{2018}).

\bibitem[{\citenamefont{Beyer et~al.}(2019)\citenamefont{Beyer, H\"olsch,
  Hussels, Cheng, Salumbides, Eikema, Ubachs, Jungen, and Merkt}}]{Beyer:19}
\bibinfo{author}{\bibfnamefont{M.}~\bibnamefont{Beyer}},
  \bibinfo{author}{\bibfnamefont{N.}~\bibnamefont{H\"olsch}},
  \bibinfo{author}{\bibfnamefont{J.}~\bibnamefont{Hussels}},
  \bibinfo{author}{\bibfnamefont{C.-F.} \bibnamefont{Cheng}},
  \bibinfo{author}{\bibfnamefont{E.~J.} \bibnamefont{Salumbides}},
  \bibinfo{author}{\bibfnamefont{K.~S.~E.} \bibnamefont{Eikema}},
  \bibinfo{author}{\bibfnamefont{W.}~\bibnamefont{Ubachs}},
  \bibinfo{author}{\bibfnamefont{C.}~\bibnamefont{Jungen}}, \bibnamefont{and}
  \bibinfo{author}{\bibfnamefont{F.}~\bibnamefont{Merkt}},
  \bibinfo{journal}{Phys. Rev. Lett.} \textbf{\bibinfo{volume}{123}},
  \bibinfo{pages}{163002} (\bibinfo{year}{2019}).

\bibitem[{\citenamefont{Fleurbaey et~al.}(2023)\citenamefont{Fleurbaey,
  Koroleva, Kassi, and Campargue}}]{Fleurbaey:23}
\bibinfo{author}{\bibfnamefont{H.}~\bibnamefont{Fleurbaey}},
  \bibinfo{author}{\bibfnamefont{A.~O.} \bibnamefont{Koroleva}},
  \bibinfo{author}{\bibfnamefont{S.}~\bibnamefont{Kassi}}, \bibnamefont{and}
  \bibinfo{author}{\bibfnamefont{A.}~\bibnamefont{Campargue}},
  \bibinfo{journal}{Phys. Chem. Chem. Phys.} \textbf{\bibinfo{volume}{25}},
  \bibinfo{pages}{14749} (\bibinfo{year}{2023}),
  \urlprefix\url{http://dx.doi.org/10.1039/D3CP01136D}.

\bibitem[{\citenamefont{Cozijn et~al.}(2023)\citenamefont{Cozijn, Diouf, and
  Ubachs}}]{Cozijn:23}
\bibinfo{author}{\bibfnamefont{F.~M.~J.} \bibnamefont{Cozijn}},
  \bibinfo{author}{\bibfnamefont{M.~L.} \bibnamefont{Diouf}}, \bibnamefont{and}
  \bibinfo{author}{\bibfnamefont{W.}~\bibnamefont{Ubachs}},
  \bibinfo{journal}{Phys. Rev. Lett.} \textbf{\bibinfo{volume}{131}},
  \bibinfo{pages}{073001} (\bibinfo{year}{2023}).

\bibitem[{\citenamefont{Lamperti et~al.}(2023)\citenamefont{Lamperti,
  Rutkowski, Ronchetti, Gatti, Gotti, Cerullo, Thibault, Jóźwiak, Wójtewicz,
  Masłowski et~al.}}]{Lamperti:23}
\bibinfo{author}{\bibfnamefont{M.}~\bibnamefont{Lamperti}},
  \bibinfo{author}{\bibfnamefont{L.}~\bibnamefont{Rutkowski}},
  \bibinfo{author}{\bibfnamefont{D.}~\bibnamefont{Ronchetti}},
  \bibinfo{author}{\bibfnamefont{D.}~\bibnamefont{Gatti}},
  \bibinfo{author}{\bibfnamefont{R.}~\bibnamefont{Gotti}},
  \bibinfo{author}{\bibfnamefont{G.}~\bibnamefont{Cerullo}},
  \bibinfo{author}{\bibfnamefont{F.}~\bibnamefont{Thibault}},
  \bibinfo{author}{\bibfnamefont{H.}~\bibnamefont{Jóźwiak}},
  \bibinfo{author}{\bibfnamefont{S.}~\bibnamefont{Wójtewicz}},
  \bibinfo{author}{\bibfnamefont{P.}~\bibnamefont{Masłowski}},
  \bibnamefont{et~al.}, \bibinfo{journal}{Commun. Phys.}
  \textbf{\bibinfo{volume}{6}}, \bibinfo{pages}{67} (\bibinfo{year}{2023}).

\bibitem[{\citenamefont{Fast and Meek}(2020)}]{Fast:20}
\bibinfo{author}{\bibfnamefont{A.}~\bibnamefont{Fast}} \bibnamefont{and}
  \bibinfo{author}{\bibfnamefont{S.~A.} \bibnamefont{Meek}},
  \bibinfo{journal}{Phys. Rev. Lett.} \textbf{\bibinfo{volume}{125}},
  \bibinfo{pages}{023001} (\bibinfo{year}{2020}).

\bibitem[{\citenamefont{{H2Spectre ver. 7.4 F}ortran~source
  code}(2022)}]{H2Spectre}
\bibinfo{author}{\bibnamefont{{H2Spectre ver. 7.4 F}ortran~source code}}
  (\bibinfo{year}{2022}),
  \bibinfo{note}{{\url{https://qcg.home.amu.edu.pl/H2Spectre.html}}},
  \urlprefix\url{{https://www.fuw.edu.pl/$\sim$krp/codes.html;
  https://qcg.home.amu.edu.pl/H2Spectre.html}}.

\bibitem[{\citenamefont{Pachucki and Komasa}(2016)}]{PK:16}
\bibinfo{author}{\bibfnamefont{K.}~\bibnamefont{Pachucki}} \bibnamefont{and}
  \bibinfo{author}{\bibfnamefont{J.}~\bibnamefont{Komasa}},
  \bibinfo{journal}{J. Chem. Phys.} \textbf{\bibinfo{volume}{144}},
  \bibinfo{eid}{164306} (\bibinfo{year}{2016}).

\bibitem[{\citenamefont{Pachucki and Komasa}(2018{\natexlab{a}})}]{PK:18a}
\bibinfo{author}{\bibfnamefont{K.}~\bibnamefont{Pachucki}} \bibnamefont{and}
  \bibinfo{author}{\bibfnamefont{J.}~\bibnamefont{Komasa}},
  \bibinfo{journal}{Phys. Chem. Chem. Phys.} \textbf{\bibinfo{volume}{20}},
  \bibinfo{pages}{247} (\bibinfo{year}{2018}{\natexlab{a}}).

\bibitem[{\citenamefont{Pachucki and Komasa}(2018{\natexlab{b}})}]{PK:18b}
\bibinfo{author}{\bibfnamefont{K.}~\bibnamefont{Pachucki}} \bibnamefont{and}
  \bibinfo{author}{\bibfnamefont{J.}~\bibnamefont{Komasa}},
  \bibinfo{journal}{Phys. Chem. Chem. Phys.} \textbf{\bibinfo{volume}{20}},
  \bibinfo{pages}{26297} (\bibinfo{year}{2018}{\natexlab{b}}).

\bibitem[{\citenamefont{Pachucki and Komasa}(2019)}]{PK:19}
\bibinfo{author}{\bibfnamefont{K.}~\bibnamefont{Pachucki}} \bibnamefont{and}
  \bibinfo{author}{\bibfnamefont{J.}~\bibnamefont{Komasa}},
  \bibinfo{journal}{Phys. Chem. Chem. Phys.} \textbf{\bibinfo{volume}{21}},
  \bibinfo{pages}{10272} (\bibinfo{year}{2019}).

\bibitem[{\citenamefont{Pachucki and Komasa}(2022)}]{PK:22}
\bibinfo{author}{\bibfnamefont{K.}~\bibnamefont{Pachucki}} \bibnamefont{and}
  \bibinfo{author}{\bibfnamefont{J.}~\bibnamefont{Komasa}},
  \bibinfo{journal}{Mol. Phys.} \textbf{\bibinfo{volume}{120}},
  \bibinfo{pages}{e2040627} (\bibinfo{year}{2022}).

\bibitem[{\citenamefont{Fromm and Hill}(1987)}]{Fromm:87}
\bibinfo{author}{\bibfnamefont{D.~M.} \bibnamefont{Fromm}} \bibnamefont{and}
  \bibinfo{author}{\bibfnamefont{R.~N.} \bibnamefont{Hill}},
  \bibinfo{journal}{Phys. Rev. A} \textbf{\bibinfo{volume}{36}},
  \bibinfo{pages}{1013} (\bibinfo{year}{1987}).

\bibitem[{\citenamefont{Harris}(2009)}]{Harris:09}
\bibinfo{author}{\bibfnamefont{F.~E.} \bibnamefont{Harris}},
  \bibinfo{journal}{Phys. Rev. A} \textbf{\bibinfo{volume}{79}},
  \bibinfo{pages}{032517} (\bibinfo{year}{2009}).

\bibitem[{\citenamefont{James and Coolidge}(1933)}]{James:33}
\bibinfo{author}{\bibfnamefont{H.~M.} \bibnamefont{James}} \bibnamefont{and}
  \bibinfo{author}{\bibfnamefont{A.~S.} \bibnamefont{Coolidge}},
  \bibinfo{journal}{J. Chem. Phys.} \textbf{\bibinfo{volume}{1}},
  \bibinfo{pages}{825} (\bibinfo{year}{1933}).

\bibitem[{\citenamefont{Pachucki}(2009)}]{Pachucki:09}
\bibinfo{author}{\bibfnamefont{K.}~\bibnamefont{Pachucki}},
  \bibinfo{journal}{Phys. Rev. A} \textbf{\bibinfo{volume}{80}},
  \bibinfo{pages}{032520} (\bibinfo{year}{2009}).

\bibitem[{\citenamefont{Pachucki}(2012)}]{Pachucki:12b}
\bibinfo{author}{\bibfnamefont{K.}~\bibnamefont{Pachucki}},
  \bibinfo{journal}{Phys. Rev. A} \textbf{\bibinfo{volume}{86}},
  \bibinfo{pages}{052514} (\bibinfo{year}{2012}).

\bibitem[{\citenamefont{Abramowitz and Stegun}(1965)}]{Abramowitz:65}
\bibinfo{author}{\bibfnamefont{M.}~\bibnamefont{Abramowitz}} \bibnamefont{and}
  \bibinfo{author}{\bibfnamefont{I.}~\bibnamefont{Stegun}},
  \emph{\bibinfo{title}{{Handbook of Mathematical Functions: With Formulas,
  Graphs, and Mathematical Tables}}}, Applied mathematics series
  (\bibinfo{publisher}{Dover Publications}, \bibinfo{year}{1965}), ISBN
  \bibinfo{isbn}{9780486612720}, \bibinfo{note}{{, Eq. (6.1.4)}}.

\bibitem[{\citenamefont{Fousse et~al.}(2007)\citenamefont{Fousse, Hanrot,
  Lef\`{e}vre, P\'{e}lissier, and Zimmermann}}]{MPFR:07}
\bibinfo{author}{\bibfnamefont{L.}~\bibnamefont{Fousse}},
  \bibinfo{author}{\bibfnamefont{G.}~\bibnamefont{Hanrot}},
  \bibinfo{author}{\bibfnamefont{V.}~\bibnamefont{Lef\`{e}vre}},
  \bibinfo{author}{\bibfnamefont{P.}~\bibnamefont{P\'{e}lissier}},
  \bibnamefont{and}
  \bibinfo{author}{\bibfnamefont{P.}~\bibnamefont{Zimmermann}},
  \bibinfo{journal}{ACM Trans. Math. Softw.} \textbf{\bibinfo{volume}{33}},
  \bibinfo{pages}{13–es} (\bibinfo{year}{2007}), ISSN
  \bibinfo{issn}{0098-3500}.

\bibitem[{\citenamefont{Bailey}()}]{MPFUN:20}
\bibinfo{author}{\bibfnamefont{D.~H.} \bibnamefont{Bailey}},
  \emph{\bibinfo{title}{{MPFUN2020}: A new thread-safe arbitrary precision
  package}},
  \bibinfo{note}{\url{https://www.davidhbailey.com/dhbpapers/mpfun2020.pdf}}.

\bibitem[{\citenamefont{Wang and Yan}(2018)}]{Wang:18a}
\bibinfo{author}{\bibfnamefont{L.~M.} \bibnamefont{Wang}} \bibnamefont{and}
  \bibinfo{author}{\bibfnamefont{Z.-C.} \bibnamefont{Yan}},
  \bibinfo{journal}{Phys. Rev. A} \textbf{\bibinfo{volume}{97}},
  \bibinfo{pages}{060501} (\bibinfo{year}{2018}).

\bibitem[{\citenamefont{Puchalski et~al.}(2018)\citenamefont{Puchalski,
  Spyszkiewicz, Komasa, and Pachucki}}]{PSKP:18}
\bibinfo{author}{\bibfnamefont{M.}~\bibnamefont{Puchalski}},
  \bibinfo{author}{\bibfnamefont{A.}~\bibnamefont{Spyszkiewicz}},
  \bibinfo{author}{\bibfnamefont{J.}~\bibnamefont{Komasa}}, \bibnamefont{and}
  \bibinfo{author}{\bibfnamefont{K.}~\bibnamefont{Pachucki}},
  \bibinfo{journal}{Phys. Rev. Lett.} \textbf{\bibinfo{volume}{121}},
  \bibinfo{pages}{073001} (\bibinfo{year}{2018}).

\bibitem[{\citenamefont{Stanke and Adamowicz}(2013)}]{Stanke:13}
\bibinfo{author}{\bibfnamefont{M.}~\bibnamefont{Stanke}} \bibnamefont{and}
  \bibinfo{author}{\bibfnamefont{L.}~\bibnamefont{Adamowicz}},
  \bibinfo{journal}{J. Phys. Chem. A} \textbf{\bibinfo{volume}{117}},
  \bibinfo{pages}{10129} (\bibinfo{year}{2013}).

\bibitem[{\citenamefont{Tiesinga et~al.}(2021)\citenamefont{Tiesinga, Mohr,
  Newell, and Taylor}}]{CODATA:18}
\bibinfo{author}{\bibfnamefont{E.}~\bibnamefont{Tiesinga}},
  \bibinfo{author}{\bibfnamefont{P.~J.} \bibnamefont{Mohr}},
  \bibinfo{author}{\bibfnamefont{D.~B.} \bibnamefont{Newell}},
  \bibnamefont{and} \bibinfo{author}{\bibfnamefont{B.~N.}
  \bibnamefont{Taylor}}, \bibinfo{journal}{Rev. Mod. Phys.}
  \textbf{\bibinfo{volume}{93}}, \bibinfo{pages}{025010}
  (\bibinfo{year}{2021}).

\bibitem[{\citenamefont{Komasa et~al.}(2019)\citenamefont{Komasa, Puchalski,
  Czachorowski, \L{}ach, and Pachucki}}]{Komasa:19}
\bibinfo{author}{\bibfnamefont{J.}~\bibnamefont{Komasa}},
  \bibinfo{author}{\bibfnamefont{M.}~\bibnamefont{Puchalski}},
  \bibinfo{author}{\bibfnamefont{P.}~\bibnamefont{Czachorowski}},
  \bibinfo{author}{\bibfnamefont{G.}~\bibnamefont{\L{}ach}}, \bibnamefont{and}
  \bibinfo{author}{\bibfnamefont{K.}~\bibnamefont{Pachucki}},
  \bibinfo{journal}{Phys. Rev. A} \textbf{\bibinfo{volume}{100}},
  \bibinfo{pages}{032519} (\bibinfo{year}{2019}).

\end{thebibliography}

\appendix

\section{Inhomogeneous terms, \texorpdfstring{$P_\beta$}{P\_beta}}
\label{App:Pbeta}

The PDE~(\ref{PDE}) is satisfied by
the general four body integral of Eq.~(\ref{Eq:gnon}). These equations involve inhomogeneous terms
$P_\beta$ with $\beta = u_i, w_i$. All these terms can be expressed by a single general function $P$
\begin{eqnarray}
P_{w_1} &=& P(w_1,u_1; w_2, u_2; w_3,u_3) \nonumber \\
       &=& P(w_1,u_1; w_3, u_3; w_2,u_2)\,, \nonumber \\
P_{u_1} &=& P(u_1,w_1; w_2, u_2; u_3,w_3)\,,\nonumber \\
P_{w_2} &=& P(w_2,u_2; w_3, u_3; w_1,u_1)\,,\nonumber \\
P_{u_2} &=& P(u_2,w_2; w_3, u_3; u_1,w_1)\,,\nonumber \\
P_{w_3} &=& P(w_3,u_3; w_1, u_1; w_2,u_2)\,,\nonumber \\
P_{u_3} &=& P(u_3,w_3; u_1, w_1; w_2,u_2)\,.
\label{A1}
\end{eqnarray}
The explicit formula for $P$ was obtained in \cite{Pachucki:12b} and is repeated here for completeness
\begin{widetext}
\begin{eqnarray}
    & &  P(w_1,u_1;w_2,u_2;w_3,u_3) \nonumber \\[0.5em]
&=& 
\frac{u_1\,w_1\,[(u_1 + w_2)^2 - u_3^2]}{(-u_1 + u_3 - w_2)\,(u_1 + u_3 + w_2)}\,
\ln\biggl[\frac{u_2 + u_3 + w_1}{u_1 + u_2 + w_1 + w_2}\biggr]
\nonumber \\[1ex] &&
+ \frac{u_1\,w_1\,[(u_1 + u_3)^2 - w_2^2]}{(-u_1 - u_3 + w_2)\,(u_1 + u_3 + w_2)}\,
\ln\biggl[\frac{w_1 + w_2 + w_3}{u_1 + u_3 + w_1 + w_3}\biggr] 
\nonumber \\[1ex] &&
 - \frac{u_1^2\,w_1^2 + u_2^2\,w_2^2 - u_3^2\,w_3^2 + w_1\,w_2\,(u_1^2 + u_2^2 - w_3^2)}{(-w_1 - w_2 + w_3)\,(w_1 + w_2 + w_3)}\,
\ln\biggl[\frac{u_1 + u_2 + w_3}{u_1 + u_2 + w_1 + w_2}\biggr] 
\nonumber \\[1ex] &&
- \frac{u_1^2\,w_1^2 - u_2^2\,w_2^2 + u_3^2\,w_3^2 + w_1\,w_3\,(u_1^2 + u_3^2 - w_2^2) }{(-w_1 + w_2 - w_3)\,(w_1 + w_2 + w_3)}\,
\ln\biggl[\frac{u_1 + u_3 + w_2}{u_1 + u_3 + w_1 + w_3}\biggr]
\nonumber \\[1ex] &&
+ \frac{u_2\,(u_2 + w_1)\,(u_1^2 + u_3^2 - w_2^2) - u_3^2\,(u_1^2 + u_2^2 - w_3^2)}{(-u_2 + u_3 - w_1)\,(u_2 + u_3 + w_1)}\,
\ln\biggl[\frac{u_1 + u_3 + w_2}{u_1 + u_2 + w_1 + w_2}\biggr]
\nonumber \\[1ex] &&
+ \frac{u_3\,(u_3 + w_1)\,(u_1^2 + u_2^2 - w_3^2) - u_2^2\,(u_1^2 + u_3^2 - w_2^2) }{(u_2 - u_3 - w_1)\,(u_2 + u_3 + w_1)}\,
\ln\biggl[\frac{u_1 + u_2 + w_3}{u_1 + u_3 + w_1 + w_3}\biggr] 
\nonumber \\[1ex] &&
- \frac{w_1\,[w_2\,(u_1^2 - u_2^2 + w_3^2) + w_3\,(u_1^2 - u_3^2 + w_2^2)]}{(w_1 - w_2 - w_3)\,(w_1 + w_2 + w_3)}\,
\ln\biggl[\frac{u_2 + u_3 + w_1}{u_2 + u_3 + w_2 + w_3}\biggr] 
\nonumber \\[1ex] &&
- \frac{w_1\,[u_2\,(u_1^2 + u_3^2 - w_2^2) + u_3\,(u_1^2 + u_2^2 - w_3^2)]}{(-u_2 - u_3 + w_1)\,(u_2 + u_3 + w_1)}\,
\ln\biggl[\frac{w_1 + w_2 + w_3}{u_2 + u_3 + w_2 + w_3}\biggr]
\label{Eq:P}.
\end{eqnarray}
\end{widetext}

In the main text, we often referred to equation~(\ref{PDE}) with $\beta$ being a linear combination
of $w_i, u_i$ as in Eq.~(\ref{Eq:tw1xyuw}). In such a case, $P_\beta$ can be obtained
from one of the following equations
\begin{align}
P_w &= P_{w_2}+P_{w_3}\,, &
P_x &= P_{w_2}-P_{w_3}\,, \nn \\
P_u &= P_{u_2}+P_{u_3}\,, &
P_y &= P_{u_3}-P_{u_2}\,.
\label{Eq:Pyxuw}
\end{align}

\section[Explicit formulas for \texorpdfstring{$G_\beta(\alpha)$}{G\_beta(alpha)}]{Explicit formulas for \texorpdfstring{$G_\beta(\alpha)$}{G\_beta(alpha)} functions and their derivatives}
\label{App:Gbeta}

The functions $G_\beta(\alpha)$ were defined in Eq.~(\ref{Eq:Gbeta}). Explicit formulas
for these functions and their derivatives with respect $\alpha$, $G_\beta'(\alpha)$,
needed for evaluation of $H_0'$ in Eq.~(\ref{Eq:H0p}), are shown below.
\begin{align}
G_{u_1}(4)&=\frac{-t + u}{t\,u\,(t + 2 u)^2}\,,\\
G'_{u_1}(4)&
=-\frac{1}{8 t u^2} + \frac{5}{4 u (t + 2 u)^2} + \frac{1}{8 u^2 (t + 2 u)}\nn\\
&\quad+\frac{\ln(2u)}{t\,u\,(t+2u)}-\frac{3\ln(t+2u)}{t\,(t+2u)^2}\,,\\
G'_{w_1}(3)&=-\frac{1}{2 u^2} -\frac{\pi^2}{48 u^2} + \frac{2}{u (t + 2 u)} - \frac{\ln(2 u)}{2 u^2}\\
&\quad+ \frac{2 \ln(t + 2 u)}{u (t + 2 u)} - \frac{\ln(\frac{t + 2 u}{2 u})}{t u}
 +\frac{\Li(\frac{t}{t+2u})}{2 u^2}\,,\nn\\
\frac{\partial G'_{w_1}(3)}{\partial t}&
=-\frac{1}{2 t u^2} + \frac{1}{2 u^2 (t + 2 u)}
 \frac{2 (t+u) \ln \left(\frac{t+2 u}{2 u}\right)}{t^2 u (t+2 u)}\nn\\
&\quad-\frac{2 \ln (t+2 u)}{u (t+2 u)^2}\,.
\end{align}

\end{document}